\documentclass[a4paper,11pt]{article}

\usepackage{graphicx, amstext, amsmath, amssymb,color}
\usepackage[colorlinks=true,linkcolor=blue]{hyperref}
\usepackage{subcaption}
\usepackage[a4paper, left=2.5cm, right=2.5cm, top=1.8cm, bottom=1.8cm, includehead, includefoot, head=30pt]{geometry}

\usepackage{pdflscape}

\title{Penalized bias reduction in extreme value estimation for censored Pareto-type data, and  long-tailed insurance applications}
\author{ J. Beirlant $^{a,b}$\footnote{Corresponding author: Jan Beirlant, KU Leuven, Dept of Mathematics, Celestijnenlaan 200B, 3001 Heverlee, Belgium; Email: jan.beirlant@kuleuven.be }, G. Maribe $^{b}$,  A. Verster$^{b}$  \\
{\fontsize{8pt}{11pt} \selectfont $^a$ Dept. of Mathematics, LStat and LRisk, KU Leuven}
\\
{\fontsize{8pt}{11pt} \selectfont $^b$ Dept. of Mathematical Statistics and Actuarial Science,  University of the Free State }
\\
}
\begin{document}

 \maketitle
\begin{abstract}
{\noindent The subject of tail estimation for randomly censored data from a heavy tailed distribution receives growing attention, motivated by applications for instance in actuarial statistics. The bias of the available estimators of the extreme value index can be substantial and depends strongly on the amount of censoring.  We review the available estimators, propose a new bias reduced estimator, and show how shrinkage estimation can help to keep the MSE under control. A bootstrap algorithm is proposed to construct confidence intervals. We compare these new  proposals with the existing estimators through simulation.  We conclude this paper with a detailed study of a long-tailed car insurance portfolio, which typically exhibit heavy censoring.
 }
\end{abstract}



\section{Introduction}  
\label{Sec1}              
Extreme value analysis under random right censoring is becoming more popular with applications for example in survival analysis, reliability and insurance. For instance, in certain long-tailed insurance products, such as car liability insurance, long developments of claims are encountered. At evaluation of the portfolio a large proportion of the  claims are then not fully developed and hence are censored.

\vspace{0.3cm}
In the setting of random right censoring the variable of interest $X$ with distribution function (df) $F$ can be censored by a random variable $C$ with df $G$. Moreover observations of $X$ and $C$ are assumed to be independent. One then  observes $Z=\min (X,C)$ with df $H$ satisfying 
$1-H=(1-F)(1-G)$, jointly with the indicator $\delta =1_{(X \leq C)}$ which equals 1 if the observation $Z$ is non-censored. Here we assume that $X$ and $C$ both are Pareto-type distributed with extreme value index (EVI) $\gamma_1 >0$ and $\gamma_2 >0$, i.e. 
\[
\bar{F} (x) = 1-F(x) = x^{-1/\gamma_1}\ell_1 (x) \mbox{ and } \bar{G}(y) = 1-G(y) = y^{-1/\gamma_2}\ell_2 (y), \;\; x,y>1,
\]
where both $\ell_1, \ell_2$ are slowly varying at infinity:
\[
\ell_j(tx)/\ell_j (t) \to_{t \to \infty} 1, \;\; \mbox{ for every } x>1 \;\; (j=1,2).
\]
Note that $1-H$ then also belongs to the domain of attraction of an extreme value distribution with positive EVI $\gamma = { {\gamma_1 \gamma_2}\over{\gamma_1+\gamma_2}}$.
Of course, the smaller $\gamma_2/\gamma_1$ the heavier the censoring will be. In long-tailed insurance applications as discussed above, the proportion of censored data can well be larger than 50\%, so that the situation $\gamma_2 < \gamma_1$ is then most relevant.

\vspace{0.3cm}
In this paper we discuss the estimation of $\gamma_1$ based on independent and identically distributed (i.i.d.) observations $(Z_i,\delta_i)$ ($i=1,\ldots,n$)  with $Z_i =\min (X_i,C_i)$ and $\delta_i=1_{(X_i\leq C_i)}$, where $(X_i,C_i)$ ($i=1,\ldots,n$) are i.i.d. random variables  from $(F,G)$. In the next section we review the available estimators for $\gamma_1$ that were published in the literature. In Section 3 we propose a new bias reduced estimator which is based on an estimator proposed by Worms and Worms (2014). Moreover we show how shrinkage estimation, as introduced in Beirlant et al. (2017) in the non-censoring case, can also be used in the censoring context. In Section 4 a parametric bootstrap algorithm is proposed in order to construct confidence intervals for $\gamma_1$. We then report on a simulation study involving all available estimators and the proposed bootstrap algorithm. Finally we make a detailed study of  a motor third party liability (MTPL) case study.

\section{A review of estimators of $\gamma_1$}
\label{Sec2}

In case there is no censoring (i.e. $\gamma_2 =\infty$ and $1-H=1-F$), the Hill (1975) estimator is the benchmark estimator for  $\gamma_1=\gamma$. Denoting the ordered $Z$ data by $Z_{1,n}\leq Z_{2,n}
\leq \ldots \leq Z_{n,n}$ this estimator is given by
\[
\hat{\gamma}_{Z,k}^H= {1 \over k}\sum_{j=1}^k \log {Z_{n-j+1,n} \over Z_{n-k,n}}.
\]
This estimator follows using maximum likelihood when approximating the distribution of the peaks $Z/t$  over a threshold $t$, given $Z> t$, by a simple Pareto distribution with density $y \mapsto \gamma^{-1} y^{-\gamma^{-1}-1}$, and taking a top order statistic $Z_{n-k,n}$ as a threshold $t$. \\
It can also be found back by estimating the functional
\begin{equation}
L_t := \mathbb{E}(\log Z - \log t|Z>t)= \int_1^{\infty} {\bar{F}(ut) \over \bar{F}(t)} {du \over u}, 
\label{MElog}
\end{equation}
which tends to the extreme value index $\gamma$ of $Z$ as $t\to \infty$. In \eqref{MElog}
$\bar{F}$ is estimated by the empirical survival function $1-\hat{F}_n$, again using $Z_{n-k,n}$ as a threshold $t$. This leads to an alternative writing of $\hat{\gamma}_{Z,k}^H$ by partial summation:
\[
\hat{\gamma}_{Z,k}^{(H)}= {1 \over k}\sum_{j=1}^k j(\log Z_{n-j+1,n} - \log Z_{n-j,n}).
\]

\vspace{0.3cm}
While both approaches yield the same estimator  in the non-censoring case this is no longer the case under random censoring. 
\begin{itemize}
\item
Beirlant et al. (2007) proposed the following estimator of $\gamma_1$ using the maximum likelihood approach:
\begin {equation}
\hat{\gamma}_{1,k}^{(H)} = {\hat{\gamma}_{Z,k}^{(H)} \over \hat{p}_k} ,
\label{Censored Hill}
\end {equation}  
with $\hat{p}_k = {1\over k} \sum_{j = 1}^k \delta_{n-j+1,n}$ the proportion of non-censored observations under the largest $k$ observations of $Z$, where $\delta_{n-j+1,n}$ denotes the $\delta$ indicator attached to $Z_{n-j+1,n}$ $(1 \leq j \leq n)$. Indeed, $\hat{\gamma}_{Z,k}^H$ estimates $\gamma$ while $\hat{p}_k$ is shown to be a consistent estimator of $p=\gamma_2/(\gamma_1+\gamma_2)$. Einmahl et al. (2008)  enhanced the asymptotic analysis of this estimator and  generalized this approach by considering any classical EVI estimator $\hat{\gamma}_{Z,k}^{(.)}$ of $\gamma$, proposing the estimators 
$
\hat{\gamma}_{1,k}^{(.)}={\hat{\gamma}_{Z,k}^{(.)}\over \hat{p}_k}$.
See also    Gomes and Oliveira (2003), Gomes and Neves (2011), and Brahimi et al. (2015) for other papers  in this spirit. 
\item
Worms and Worms (2014) essentially used the second approach estimating \eqref{MElog} by substituting $1-F$ with the Kaplan-Meier estimator $1-\hat{F}^{KM}_n (x) = \Pi_{Z_{i,n} \leq x} \left(1- {1 \over n-i+1} \right)^{\delta_{i,n}}$,  setting $1-\hat{F}^{KM}_n (Z_{n,n})=0$:  
\begin {equation}
\hat{\gamma}_{1,k}^{(W)} = \sum_{j=1}^k {1-\hat{F}^{KM}_n (Z_{n-j+1,n}) \over 1-\hat{F}^{KM}_n (Z_{n-k,n})} \left( \log Z_{n-j+1,n}-\log Z_{n-j,n}\right).
\label{Worms}
\end {equation}
Worms and Worms (2014) also introduced 
\begin {equation}
\hat{\gamma}_{1,k}^{(KM)} = \sum_{j=1}^k {1-\hat{F}^{KM}_n (Z_{n-j+1,n}) \over 1-\hat{F}^{KM}_n (Z_{n-k,n})}{\delta_{n-j+1,n} \over j} \left( \log Z_{n-j+1,n}-\log Z_{n-k,n}\right).
\label{WormsKM}
\end {equation}
Through simulations the estimator $\hat{\gamma}^{(W)}_{1,k}$ was found to have the best RMSE behaviour for smaller values of $k$. Below we will then concentrate on $\hat{\gamma}_{1,k}^{(W)}$. Unfortunately, the asymptotic distribution of $\hat{\gamma}_{1,k}^{(W)}$ is not known up to now.   The asymptotic normality of a fixed threshold version of $\hat{\gamma}_{1,k}^{(KM)}$  can be derived from Worms and Worms (2017) in case $\gamma_1 < \gamma_2$. Brahimi et al. (2016) consider closely related estimators but also considered asymptotic results in case  $\gamma_1 < \gamma_2$ or $p >1/2$.  
\item
In an objective Bayesian approach (see Zellner, 1971),  Ameraoui et al. (2016) recently proposed several other estimators. They considered the maximum posterior (m) and the mean posterior  (e) estimators of the posterior density of $\gamma_1$. The maximal data information (M) prior and a conjugate gamma prior with parameters $(a,b)$ were considered. It was also shown that Jeffreys prior lead to special cases of the conjugate prior estimators setting $a=b=0$.  This then leads to the following estimators:
\begin{eqnarray}
\hat{\gamma}_{1,k}^{(m,M)}&=&{2 k\hat{\gamma}_{Z,k}^{(H)} \over 1+k\hat{p}_k+\sqrt{(1+k\hat{p}_k)^2+4k\hat{\gamma}_{Z,k}^{(H)}}} , \label{Bayes1} \\
\hat{\gamma}_{1,k}^{(e,a,b)} &=& {k\hat{\gamma}_{Z,k}^{(H)}+b \over k\hat{p}_k +a},
\label{Bayes2} \\
\hat{\gamma}_{1,k}^{(m,a,b)} &=& {k\hat{\gamma}_{Z,k}^{(H)}+b \over k\hat{p}_k +a-1}.
\label{Bayes3}
\end{eqnarray}
\end{itemize}

\vspace{0.3cm}
It is well-known that extreme value estimators often suffer from severe bias. In the random censoring case Einmahl et al.  (2008) first derived the asymptotic bias of $\hat{\gamma}_{1,k}^{(H)}$, which was further detailed in Beirlant et al. (2016) under more specific assumptions on the slowly varying functions $\ell_1 $ and $\ell_2$, which are commonly proposed in extreme value statistics:
\begin{eqnarray*}
1-F(x)&=&C_1 {x^{-1/\gamma_1}}(1+D_{1} x^{-\beta_1} (1+o(1))),\; x \to \infty,\\
1-G(y)&=&C_2 {y^{-1/\gamma_2}}(1+D_{2} x^{-\beta_2} (1+o(1))),\;  y \to \infty,
\end{eqnarray*}
where $\beta_1,\beta_2,C_1,C_2$ are positive constants and $D_1,D_2$ are real constants. Taking the bias of the Hill estimator $\hat{\gamma}^{(H)}_{Z,k}$ as a reference, it was observed  that especially when $\beta_1 \leq \beta_2$ the bias of  $\hat{\gamma}_{1,k}^{(H)}$ increases with decreasing value of $p$, i.e. for smaller $\gamma_2/\gamma_1$.
Within the maximum likelihood approach, a bias reduced estimator was then proposed for  the censoring case following the technique from Beirlant et al. (2009) where the distribution of the excesses $X/t|X>t$ is approximated by the extended Pareto (EP) distribution with df
$\mathbb{P}(Y \leq y)=1-(y\{1+\kappa_t (1-y^{-\beta_1})\})^{-1/\gamma_1}$ where $\kappa_t = \gamma_1 D_1 t^{-\beta_1} (1+o(1))$ as $t \to \infty$.  The resulting estimator is given by
\begin{equation}
\hat{\gamma}_{1,k}^{(EP)}
=\hat{\gamma}_{1,k}^{(H)}+C_{\hat{\gamma}_{1,k}^{(H)},\beta_*}{H_{Z,k}^{(-\beta_*)}\over \hat{p}_k}
\left\{
H_{Z,k}^{(-\beta_*)}-\hat{\gamma}_{1,k}^{(H)}E_{Z,k}^{(c)}(-\beta_*)
\right\}
\label{biasreduced}
\end{equation}
where, $\beta_{*}=\min(\beta_1,\beta_2)$ and   
\begin{eqnarray*}
H_{Z,k}^{(-\beta_*)}&=&{1\over\beta_*}
\left(1-{1\over{k}}\sum_{j=1}^k \left({Z_{n-j+1,n}\over Z_{n-k,n}}\right)^{-\beta_*}\right), \\
E_{Z,k}^{(c)}(-\beta_*)&=&{1\over{k}}\sum_{j=1}^k \delta_{n-j+1,n}{\left(Z_{n-j+1,n}\over{Z_{n-k,n}}\right)}^{-\beta_*},\\ C_{\gamma,\beta_*}&=&-{{(1+\gamma\beta_*)^3(1+2\gamma\beta_*)}\over{\gamma^4\beta_*^3}}.
\end{eqnarray*}
In this estimation procedure $\beta_*$ is assumed to be known. In fact, since in the definition of the EP distribution the term $y^{-\beta_*}$ is multiplied by the $\kappa_t$-factor, the asymptotic distribution of tail estimators  based on the EP distribution will not depend on the asymptotic distribution of an estimator of $\beta_*$. One can also impute estimators of the parameter $\beta_*$ of the distribution $H$ of $Z$ without increasing the bias  in estimating $\gamma_1$. Estimators of $\rho_* = -\gamma \, \beta_*$ were discussed in Fraga Alves {\it et al.} (2003). An estimator for $\beta_*$ is then given by $-\rho_*/ \hat{\gamma}_{Z,k}^{(H)}$. In the simulations the sensitivity of the choice of $\rho_*$ was examined.
In \eqref{biasreduced} one can reparametrize $\beta_* \hat{\gamma}_{Z,k}^{(H)}$ by $-\rho_*$ with $\rho_* <0$, leading to
\begin{equation}
\hat{\gamma}_{1,k}^{(EP)}(\rho_*)=
\hat{\gamma}_{1,k}^{(H)}- 
{\hat{\gamma}_{1,k}^{(H)} (1-\rho_*)^2(1-2\rho_*)\over \rho_*^3}
\left\{
H_{Z,k}^{(\rho_*/ \hat{\gamma}_{Z,k}^{(H)})}-\hat{\gamma}_{1,k}^{(H)}E_{Z,k}^{(c)}(\rho_*/ \hat{\gamma}_{Z,k}^{(H)})
 \right\}.
\label{EPrho}
\end{equation}
It was  shown that when using the correct value of $\beta_*$ or $\rho_*$, the asymptotic bias of $\hat{\gamma}_{1,k}^{(EP)}$ is 0 as long as $\sqrt{k} (k/n)^{\beta_*} = O(1)$, whereas the asymptotic bias of the original estimator $\hat{\gamma}^{(H)}_{1,k}$ is only 0 when $\sqrt{k} (k/n)^{\beta_*} \to 0$ as $k,n \to \infty$. Hence the bias is reduced for a longer set of values of $k \geq 1$ when choosing $t$ as $X_{n-k,n}$. At the other hand the variance of this bias reduced estimator was shown to be increased by the factor $\left({1+\gamma\beta_*\over \gamma \beta_*} \right)^2$ in comparison with the estimator $\hat{\gamma}^{(H)}_{1,k}$.

\vspace{0.3cm}
Before comparing the different estimators through simulations, we next derive  a bias reduced estimator starting from the Worms \& Worms estimator $\hat{\gamma}_{1,k}^{(W)}$ from \eqref{Worms}. Following Beirlant et al. (2017), we then also apply  shrinkage estimation on $\kappa_t$ forcing this parameter to decrease to $0$ as $t \to \infty$ or $k \downarrow 1$ as it is the case in the mathematical definition of $\kappa_t$.

\section{Bias reduction of the Worms \& Worms estimator and penalized estimation of bias }  
\label{Sec3}

Using the EP approximation to the survival function ${\bar{F}(ut) \over \bar{F}(t)} $ of the excesses $X/t|X>t$, leads to the following approximation of the integral expression of $L_t$ in \eqref{MElog}: as $t\to \infty$
\begin{equation}
L_t = \int_1^{\infty} {\bar{F}(ut) \over \bar{F}(t)} {du \over u} = \gamma_1 - \kappa_t {\beta_1 \gamma_1 \over 1+\beta_1 \gamma_1}(1+o(1)).
\label{Lt}
\end{equation}
Similarly, considering 
$$E_t (-\beta_1):= \mathbb{E} \left( ({X \over t})^{-\beta_1}|X>t \right) = 1+ \int_1^{\infty}{\bar{F}(ut) \over \bar{F}(t)}
du^{-\beta_1}$$ leads to 
\begin{equation}
(1+\gamma_1\beta_1) E_t (-\beta_1) = 1 + {\kappa_t \over \gamma_1} {(\beta_1 \gamma_1)^2 \over 1+2\beta_1 \gamma_1}(1+o(1)).
\label{Et}
\end{equation}
Substituting $\gamma_1$ in the left hand side of \eqref{Et} by
the expression $L_t + \kappa_t {\beta_1 \gamma_1 \over 1+\beta_1 \gamma_1}$ which follows from \eqref{Lt}, one obtains for $t \to 0$ that
\begin{eqnarray*}
&& \hspace{-1cm}(1+\beta_1 L_t) E_t(-\beta_1) \\&=& 1+ \left\{ {\kappa_t \over \gamma_1}
{(\beta_1 \gamma_1)^2 \over 1+2\beta_1 \gamma_1}- \kappa_t E_t(-\beta_1){\beta_1^2 \gamma_1 \over 1+\beta_1 \gamma_1}E_t(-\beta_1)\right\} (1+o(1))\\
&=& 1+ {\kappa_t \over \gamma_1}(\beta_1 \gamma_1)^2 [{1 \over 1+2\beta_1\gamma_1}-{1 \over (1+\beta_1\gamma_1)^2}](1+o(1))\\
&=&  1+ {\kappa_t \over \gamma_1}{ (\beta_1 \gamma_1)^4 \over (1+\beta_1\gamma_1)^2(1+2\beta_1\gamma_1)}(1+o(1)),
\end{eqnarray*} 
where in the second step we approximated  $E_t(-\beta_1)$  by $(1+\gamma_1\beta_1)^{-1}$. We now conclude that 
\begin{equation}
\kappa_t = {L_t (1+\beta_1 L_t)^3(1+2\beta_1 L_t) \over (\beta_1 L_t)^4}\left\{E_t(-\beta_1)-{1 \over 1+\beta_1 L_t} \right\}(1+o(1)).
\label{deltatilde}
\end{equation}
Estimating $L_t$ at a random threshold $Z_{n-k,n}$ by $\hat{\gamma}_{1,k}^{(W)}$ and similarly $E_t(-\beta_1)$  by
\[
\hat{E}_k (-\beta_1)= 1+ \sum_{j=1}^k {1-\hat{F}^{KM}_n (Z_{n-j+1,n}) \over 1-\hat{F}^{KM}_n (Z_{n-k,n})} 
\left( Z^{-\beta_1}_{n-j+1,n}- Z_{n-j,n}^{-\beta_1} \right)/Z^{-\beta_1}_{n-k,n},
\]
we obtain the following bias reduced estimator for $\gamma_1$ combining \eqref{Lt} and \eqref{deltatilde}:
\begin{equation}
\hat{\gamma}_{1,k}^{(BR,W)}=
\hat{\gamma}_{1,k}^{(W)}+ 
{\hat{\gamma}_{1,k}^{(W)} (1+\beta_1\hat{\gamma}_{1,k}^{(W)})^2(1+2\beta_1 \hat{\gamma}_{1,k}^{(W)})\over (\beta_1\hat{\gamma}_{1,k}^{(W)})^3}
\left\{\hat{E}_k (-\beta_1)-{1 \over 1+\beta_1 \hat{\gamma}_{1,k}^{(W)}} \right\}.
\label{BRW}
\end{equation}
In \eqref{BRW} one can reparametrize $\beta_1 \hat{\gamma}_{1,k}^{(W)}$ by $-\rho_1$ with $\rho_1 <0$, leading to
\begin{equation}
\hat{\gamma}_{1,k}^{(BR,W)}(\rho_1)=
\hat{\gamma}_{1,k}^{(W)}- 
{\hat{\gamma}_{1,k}^{(W)} (1-\rho_1)^2(1-2\rho_1)\over \rho_1^3}
\left\{\hat{E}_k (\rho_1/\hat{\gamma}_{1,k}^{(W)})-{1 \over 1-\rho_1} \right\},
\label{BRWrho}
\end{equation}
and we will study the sensitivity of the estimator with respect to the choice of $\rho_1$. In fact our objective will be to look for an appropriate choice of $\rho_1$ such that the plot of the estimates as a function of $k$ is most constant in order to assist practitioners.

\vspace{0.3cm}
Also here the  variance of the bias reduced estimator can be expected to be inflated compared with the corresponding estimator (here $\hat{\gamma}_{1,k}^{(W)}$).
This will be confirmed by the simulations in the next section.
However, Beirlant et al. (2017) showed that this problem can be alleviated forcing the bias estimator ${\hat{\gamma}_{1,k}^{(W)} (1-\rho_1)^2(1-2\rho_1)\over (-\rho_1)^3}
\left\{\hat{E}_k (\rho_1/\hat{\gamma}_{1,k}^{(W)})-{1 \over 1-\rho_1} \right\}$ to decrease to 0 as $t \to \infty$ or $k \downarrow 1$. Formally applying the shrinkage procedure from 
Beirlant et al. (2017) leads to a penalized version of \eqref{deltatilde}:
\[
\kappa_t^s = {1+\beta_1 L_t \over \frac{\omega L_t}{k\sigma^2_{1,k,n}} +\frac{(\beta_1 L_t)^4}{L_t (1+\beta_1 L_t)^2(1+2\beta_1 L_t)}} \left\{ E_t(-\beta_1)-{1 \over 1+\beta_1 L_t} \right\},
\]
where $\sigma^2_{1,k,n}= (k/n)^{-2\rho_1}$ and $\omega$ is a  weight factor that allows to control the penalization. The term $(\omega L_t)/(k\sigma^2_{1,k,n})$ makes the bias correction shrink for smaller values of $k$, i.e. when the original estimator $\hat{\gamma}_{1,k}^{(W)}$ is asymptotically unbiased, namely 
$k\sigma^2_{1,k,n} \to 0$. This then leads to the penalized estimator
\begin{equation}
\hat{\gamma}_{1,k}^{(s,W)}(\rho_1) = \hat{\gamma}_{1,k}^{(W)}
- {\rho_1 \over \frac{\omega \hat{\gamma}_{1,k}^{(W)} }{k\sigma^2_{1,k,n}} +\frac{\rho_1^4}{\hat{\gamma}_{1,k}^{(W)} (1-\rho_1)^2(1-2\rho_1)} }
\left\{\hat{E}_k (\rho_1/\hat{\gamma}_{1,k}^{(W)})-{1 \over 1-\rho_1} \right\}.
\label{pW}
\end{equation}
In a similar way the bias component in $\hat{\gamma}_{1,k}^{(EP)}(\rho_*)$ can be penalized for smaller values of $k$:
\begin{equation}
\hat{\gamma}_{1,k}^{(s,EP)}(\rho_*) =
\hat{\gamma}_{1,k}^{(H)}- 
{\rho_* \over \frac{\omega \hat{\gamma}_{1,k}^{(H)} }{k\sigma^2_{*,k,n}} +\frac{\rho_*^4}{\hat{\gamma}_{1,k}^{(H)} (1-\rho_*)^2(1-2\rho_*)} }
\left\{
H_{Z,k}^{(\rho_*/ \hat{\gamma}_{Z,k}^{(H)})}-\hat{\gamma}_{1,k}^{(H)}E_{Z,k}^{(c)}(\rho_*/ \hat{\gamma}_{Z,k}^{(H)})
 \right\}
\label{pEP}
\end{equation}
with $\sigma^2_{*,k,n}=(k/n)^{-2\rho_*}$.

\section{Bootstrap confidence intervals for $\gamma_1$}
\label{SecB}

Given the lack of any  distribution theory for the Worms and Worms estimator $\hat{\gamma}_{1,k}^{(W)}$ we here present a parametric bootstrap algorithm in order to construct confidence intervals for $\gamma_1$. 

\vspace{0.3cm} The main idea behind this bootstrap procedure is that for a value of $k$ where the bias of an estimator of $\gamma_1$ is 0, one can as well simulate from  simple Pareto distributions rather than from the true Pareto-type distribution $F$ and $G$ in order to construct samples of estimators. Also note that 
$$
\hat{\gamma}_{2,k}^{(W)} = \sum_{j=1}^k {1-\hat{G}^{KM}_n (Z_{n-j+1,n}) \over 1-\hat{G}^{KM}_n (Z_{n-k,n})} \left( \log Z_{n-j+1,n}-\log Z_{n-j,n}\right), 
$$
where $1-\hat{G}^{KM}_n (y)= \Pi_{Z_{i,n} \leq y} \left(1- {1 \over n-i+1} \right)^{1-\delta_{i,n}}$ denotes the Kaplan-Meier estimator of $G$, jointly with its bias reduced versions constructed in a similar way as in the preceding section (replacing $\delta_{n-j+1,n}$ by $1-\delta_{n-j+1,n}$), lead to estimates of $\gamma_2$. The procedure then runs as follows:
\begin{itemize} 
\item
Given a value of $\hat{k}_1$, respectively $\hat{k}_2$, where the bias of the estimator $\hat{\gamma}_{1,k}^{(W)}$, respectively $\hat{\gamma}_{2,k}^{(W)}$, is judged to be negligible,
one can perform a parametric bootstrap using  samples of size $n$ from 
$\left( \min(\hat{X}_i, \hat{C}_i),1_{(\hat{X}_i \leq \hat{C}_i)}\right)$ ($i=1,\ldots,n$) where $\hat{X}_i$, respectively $\hat{C}_i$, are simulated from a standard Pareto distribution with survival function $x^{-1/\hat{\gamma}_{1,\hat{k}_1}^{(W)}}, \, x>1$, respectively  $y^{-1/\hat{\gamma}_{2,\hat{k}_2}^{(W)}}, \,y>1$. 
\item
The values  $\hat{k}_j$, $j=1,2$, are chosen from 
$$
\hat{k}_j = \max \{k: |\hat{\gamma}_{j,k}^{(W)}-\hat{\gamma}_{j,k}^{(s,W)}(\rho_j)| \leq \epsilon \}
$$ for a small value of $\epsilon$.
\item
From each bootstrap sample one then retains a  bootstrap estimate $\hat{\gamma}_{1,\hat{k}_1}^{(*,s,W)}$ of $\gamma_1$. 
\item
Finally, repeating this bootstrap sampling step $N$ times, we consider the empirical distribution of the values  $\hat{\gamma}_{1,\hat{k}_1}^{(*,s,W)}(j)$ $(j=1,\ldots,N)$, and more specifically the $\lfloor N\alpha/2 \rfloor$, $\lfloor N(1-\alpha)/2 \rfloor$ empirical quantiles , in order to construct a $100(1-\alpha)\%$ confidence interval for $\gamma_1$.
\end{itemize}
In order to test this bootstrap procedure in the next section we will apply this procedure to several simulated censored samples under different values of the proportion of non-censoring.

\section{Finite sample simulations}  
\label{Sec4}

As the asymptotic distribution of $\hat{\gamma}_{1,k}^{(W)}$ and hence also of $\hat{\gamma}_{1,k}^{(BR,W)}(\rho_1)$ and $\hat{\gamma}_{1,k}^{(s,W)}(\rho_1)$ is not known, we here consider a comparison using finite sample simulations. 
 We report the simulation results for sample size $n=500$ from
\begin{itemize}
\item the Burr ($\eta,\tau,\lambda$) distribution with right tail function 
$$
1-F(x) = \left({\eta \over \eta + x^{\tau}} \right)^{\lambda}, \; x>0,
$$
with $\eta,\tau,\lambda >0$, and $\gamma = 1/(\tau\lambda), \beta = \tau, D= -\lambda \eta$;
\item the Fr\'echet ($\alpha$) distribution with right tail function 
$$
1-F(x) = 1-\exp (-x^{-\alpha}), \; x>0,
$$
with $\alpha >0$, and $\gamma = 1/\alpha, \beta = \alpha, D= -1/2$.
\end{itemize}
Here we present results concerning the bias and the root mean squared error of the different estimators of $\gamma_1$ discussed above, and of the bootstrap algorithm, in case 
\begin{itemize}
\item Burr ($10,2,2$) censored by Burr ($10,5,2$) with $\gamma_1=0.25$ and $\gamma_2=0.10$, leading to heavy censoring with the proportion of non-censoring $p=0.286$; see Figure 1;
\item Burr ($10,2,1$) censored by Burr ($10,2,1$) with $\gamma_1=\gamma_2=0.5$ so that $p=0.5$; see Figure 2;
\item Burr ($10,5,2$) censored by Burr ($10,2,2$) with $\gamma_1=0.10$ and $\gamma_2=0.25$, with light censoring $p=0.714$; see Figure 3;
\item Fr\'echet (2) censored by Fr\'echet (1) with $\gamma_1=0.5$ and $\gamma_2=1$, so that $p=2/3$; see Figure 4.
\end{itemize} 
In each of these four cases  we consider the results for
\begin{itemize}
\item $\hat{\gamma}_{1,k}^{(H)}$ from \eqref{Censored Hill}  , $\hat{\gamma}_{1,k}^{(EP)}(\rho_*)$ from \eqref{EPrho}, and $\hat{\gamma}_{1,k}^{(s,EP)}(\rho_*)$ from \eqref{pEP} with $\omega=1$ and  for different values of $\rho_*$ (left in Figures 1-4),
\item $\hat{\gamma}_{1,k}^{(W)}$ from \eqref{Worms}, $\hat{\gamma}_{1,k}^{(KM)}$ from \eqref{WormsKM}, $\hat{\gamma}_{1,k}^{(BR,W)}(\rho_1)$ from \eqref{BRWrho}, and $\hat{\gamma}_{1,k}^{(s,W)}(\rho_1)$ from \eqref{pW} with $\omega=1$ and for different values of $\rho_1$ (see middle of Figures 1-4),
\item the Bayesian estimators $\hat{\gamma}_{1,k}^{(m,M)}$,  $\hat{\gamma}_{1,k}^{(e,1,2)}$ and $\hat{\gamma}_{1,k}^{(m,1,2)}$ from \eqref{Bayes1}, \eqref{Bayes2} and \eqref{Bayes3} (right in Figures 1-4).
\end{itemize}

\vspace{0.3cm}
One observes that in case $p<0.5$ the likelihood based estimator $\hat{\gamma}_{1,k}^{(H)}$ and its bias reduced versions, and the Bayesian estimators  have larger bias than the estimators derived from estimating the functional form $L_t$ in \eqref{MElog}. Bias reduction of $\hat{\gamma}_{1,k}^{(H)}$ helps only partially in such cases. 
In these cases the bias reduced and penalized estimators $\hat{\gamma}_{1,k}^{(BR,W)}$ and $\hat{\gamma}_{1,k}^{(s,W)}(\rho_1)$  perform good with low bias and low RMSE for a long interval of values of $k$ which is quite helpful in choosing an appropriate value of $k$. This is in contrast with the bias of  $\hat{\gamma}_{1,k}^{(W)}$ and $\hat{\gamma}_{1,k}^{(KM)}$ which is systematically decreasing with decreasing value of $k$. In order to evaluate the effect of the penalization in $\hat{\gamma}_{1,k}^{(s,W)}(\rho_1)$,  we focused the scale for the bias  and RMSE plots in the Burr cases, see Figure 5. Especially in case  $p \leq 0.5$ and for smaller values of $k$ the mean of the shrinkage estimator is much more stable and ultimately for small $k$ the bahaviour of this estimator follows that of $\hat{\gamma}_{1,k}^{(W)}$. The resulting RMSE is then also  a lower envelope of the RMSE curves of $\hat{\gamma}_{1,k}^{(W)}$ and   $\hat{\gamma}_{1,k}^{(BR,W)}$.

\vspace{0.3cm}
On the other hand when $p>0.5$, and especially in the Fr\'echet case,  the basic estimators $\hat{\gamma}_{1,k}^{(H)}$, $\hat{\gamma}_{1,k}^{(W)}$ and $\hat{\gamma}_{1,k}^{(m,M)}$ work almost equally well, and this also holds for the  different approaches to bias reduction. Within the group of Bayes estimators  clearly  $\hat{\gamma}_{1,k}^{(m,M)}$ works best.

\vspace{0.3cm}
We also tested the proposed bootstrap procedure in the same cases as considered in Figures 1 to 4. We applied the algorithm with  $\alpha=0.05$ and $N=1000$  to 1000 samples of size $n=500$. In Figure 6 the 1000 confidence intervals are given when choosing $\hat{k}_1$ and $\hat{k}_2$ adaptively using $\epsilon=0.01$, and when keeping $\hat{k}_1=\hat{k}_2$ fixed to $25 = 0.05 \times n$ throughout (this value of $k$ appears  appropriate  on the basis of Figures 1 to 4). 
Further simulations showed that for $n=1000$ keeping $\hat{k}_1=\hat{k}_2$ fixed to $0.04 \times n$ leads to confidence intervals that attain the required 95\% level closely.
The confidence intervals missing the correct value of $\gamma_1$ are put in dark grey. Of course when the $k$ values are chosen adaptively,  it is more difficult to attain the confidence level $1-\alpha = 0.95$.

A deeper understanding of the distribution of $\hat{\gamma}_{1,k}^{(W)}$ and $\hat{E}_k (-\beta_1)$ appears necessary to enhance the adaptive choice of $k_1, k_2$ and the performance of the bootstrap algorithm. 

\section{A case study from car insurance}  
\label{Sec5}

Finally, in order to illustrate the merits of the newly proposed method, we consider a data set with indexed total payments from a motor third party liability insurance company operating in the EU, with records from 1995 till 2010 with  $n=849$ claims of which only 340 were completely developed at the end of 2010. For every claim the indexed cumulative payments are given at the end of every year until development. In Figure 6 we plotted the proportions of non-censored data $\hat{p}_k$ which are situated in the top $100k/n$\% of cumulative payments  at the end of 2010 as a function of $k/n$. 
In practice most companies substitute the censored observations by ultimate predictions obtained through reserving techniques. Here we show how the extreme value methods for censored data can also be used directly without ultimates in order to obtain relevant extreme value predictions.

\vspace{0.3cm}
Due to the long-tail nature of such portfolios, only the claims with arrival year between 1995 and 1999 can be considered to satisfy the condition of weak censoring $p > 0.5$ or $\gamma_1 < \gamma_2$ when using the information up to 2010. For this group of early claims 29\% is censored at the end of 2010, while the percentage of censoring is 60\% when considering all claims. Note also that when considering all claims  the largest 20 \% are all censored, whereas this amount increases to 40\% for the claims arriving after 2003. Needless to say that extreme value methodology is quite challenging in such a case.

\vspace{0.3cm}
In order to illustrate  the stability of the proposed bias reduction technique based on the Worms and Worms estimator over different percentages of censoring, we also split the full data set in groups along the arrival times in 1995-1999, 1998-2002, 2001-2005, 2004-2008, 2007-2010. In Figure 7 the original estimates $\hat{\gamma}_{1,k}^{(W)}$ and $\hat{\gamma}_{1,k}^{(BR,W)}(-3)$ are given as a function of $k/n$ for each of these subgroups and for the complete data set. The value $\rho_1=-3$ was chosen as this value yields the most constant plots as a function of $k$. The plots of $\hat{\gamma}_{1,k}^{(W)}$ are steepest for the claims which are most recent in 2010. The stability of the bias reduced estimates over these subgroups is quite convincing leading to estimates of $\gamma_1$ between 0.6 and 0.7.

\vspace{0.3cm}
As another validity check, in Figure 8 we consider only the claims from 1995-1999 with their  cumulative payments as of 2000 till 2010 in steps of 2 years. Note that in 2000 only 9\% of those claims were fully developed, while at 2010 this percentage rose to 71\%.  Again the bias reduced estimates  
$\hat{\gamma}_{1,k}^{(BR,W)}(-3)$ are remarkably stable over $k$. 

\vspace{0.3cm}
We also applied the bootstrap algorithm to this case study. 
Using $\epsilon = 0.01$ leads to $\hat{k}_1=73$, $\hat{k}_2=50$, $\hat{\gamma}_{1,73}^{(W)}=0.725$, and $\hat{\gamma}_{2,50}^{(W)}=0.652$, see Figure 10 (left). The confidence intervals for the different values of $k$ are given in Figure 10 (right) with special attention for the case $k=73$ which leads to the interval 95\% confidence interval $(0.48;0.91)$.
Choosing $\hat{k}_1=\hat{k}_2$ fixed at 4 to 5\% of the sample size $n=849$, as suggested in the simulation section, leads to lower bounds that are somewhat lower than 0.48, as can be seen from Figure 10 (right).

\section{Conclusion}  
\label{Sec6}

The estimator $\hat{\gamma}_{1,k}^{(W)}$ from Worms and Worms (2014) has the best RMSE behaviour between all available first order estimators of $\gamma_1$. In order to enhance the practical use of this estimator we proposed  bias reduction and penalization  techniques which lead to improved bias and RMSE behaviour. Moreover a bootstrap procedure is proposed in order to construct confidence intervals. This is especially useful with long-tailed insurance products.  In order to enhance the adaptive choice of the number of extreme data $k$ 	asymptotic representations of the estimators involved are needed for all cases, but especially in case of heavy censoring. This will be the subject of future work.  

\section{Acknowledgments}  
\label{Sec7}
	\noindent The authors are indebted to Rym and Julien Worms for helpful discussions and suggestions on this topic. This work is based on the research supported wholly/in part by the National Research Foundation of South Africa (Grant Number 102628). The Grantholder acknowledges that opinions, findings and conclusions or recommendations expressed in any publication generated by the NRF supported research is that of the author(s), and that the NRF accepts no liability whatsoever in this regard. 


\begin{landscape}
\begin{figure}[h]
	\centering
	\begin{subfigure}[h]{0.3\linewidth}
		\includegraphics[width=8.5cm,height=5.5cm]{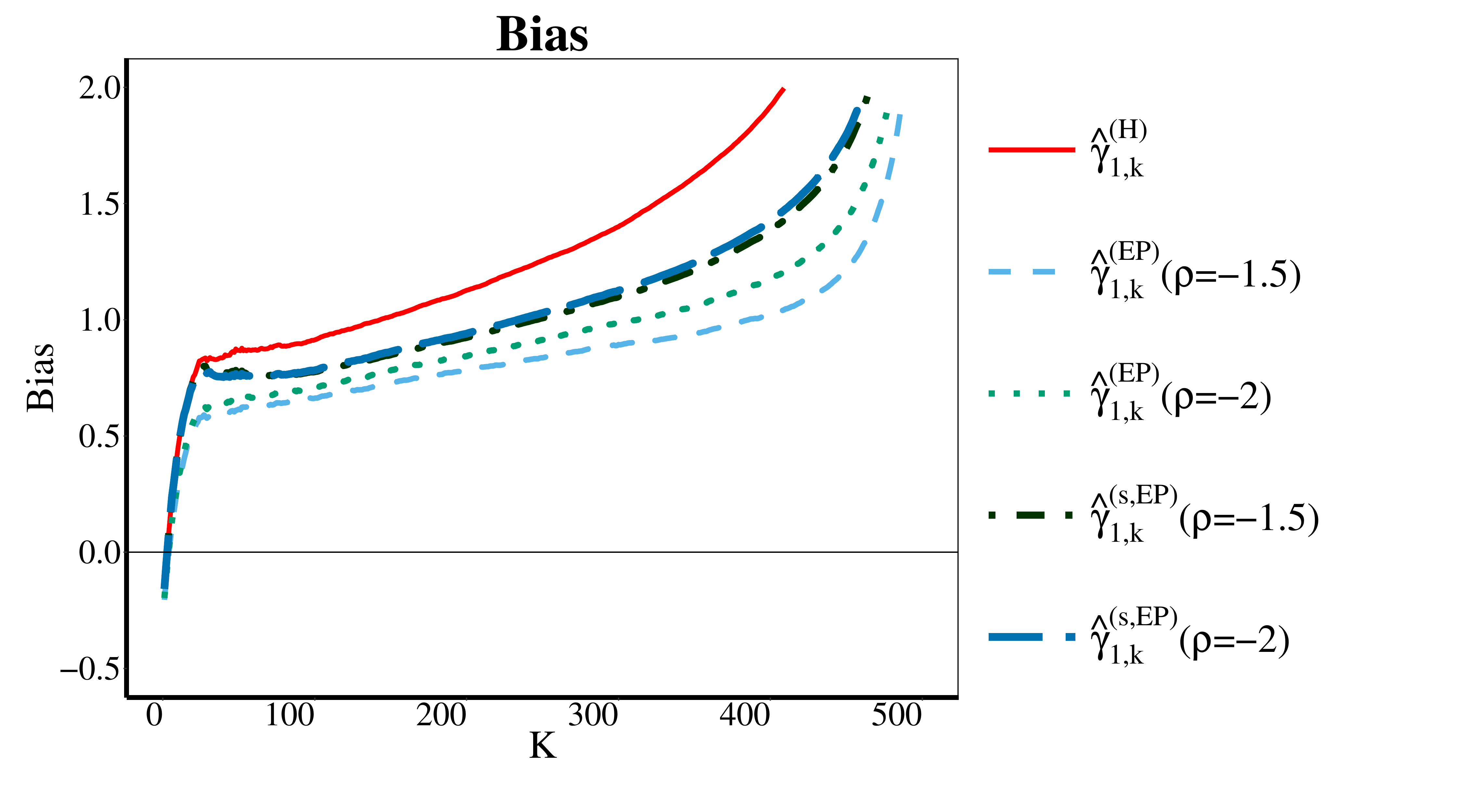}
	\end{subfigure}
	\hspace{\fill}
	\begin{subfigure}[h]{0.3\linewidth}
		\includegraphics[width=8.5cm,height=5.5cm]{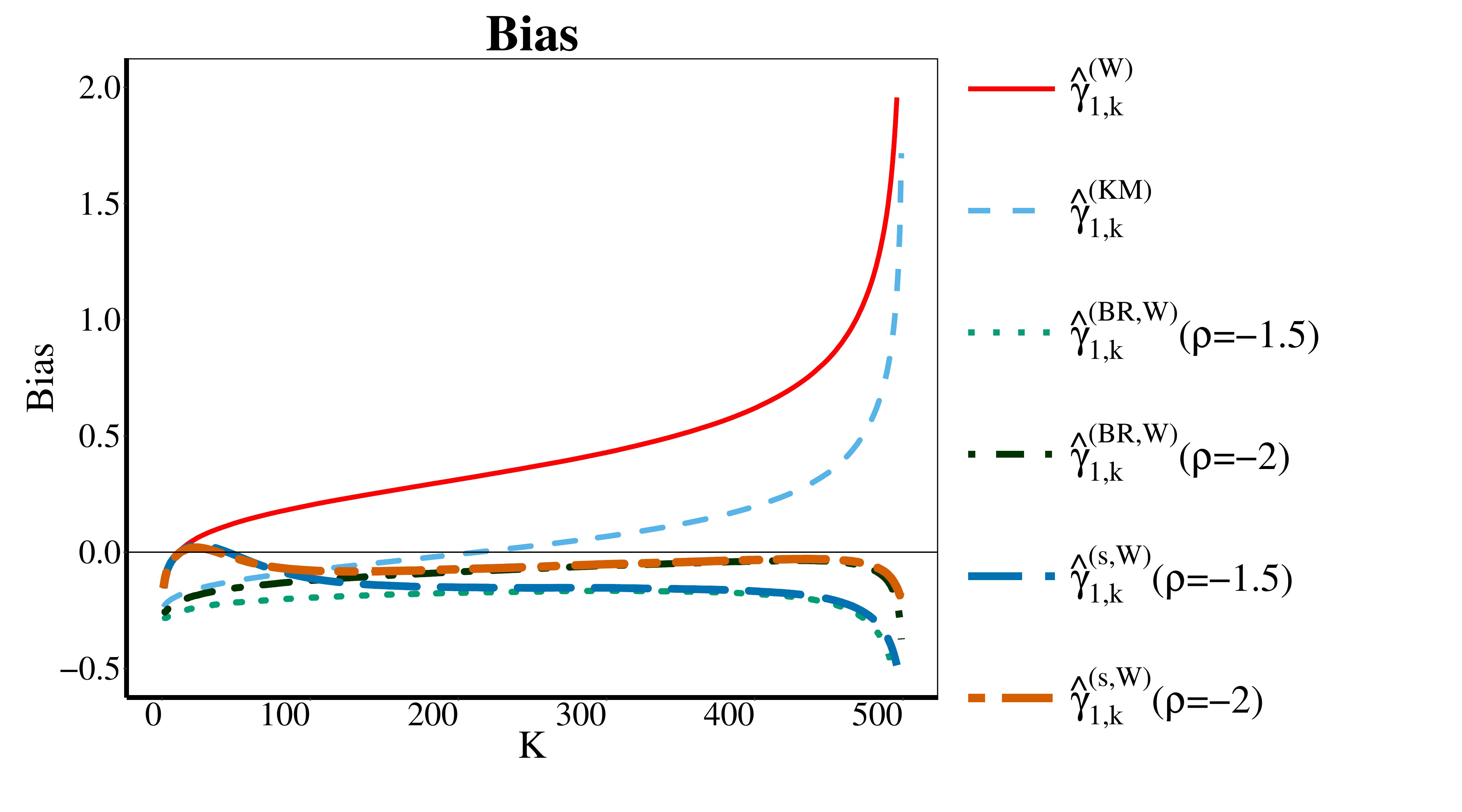}
	\end{subfigure}
	\hspace{\fill}
	\begin{subfigure}[h]{0.3\linewidth}
		\includegraphics[width=7.8cm,height=5.5cm]{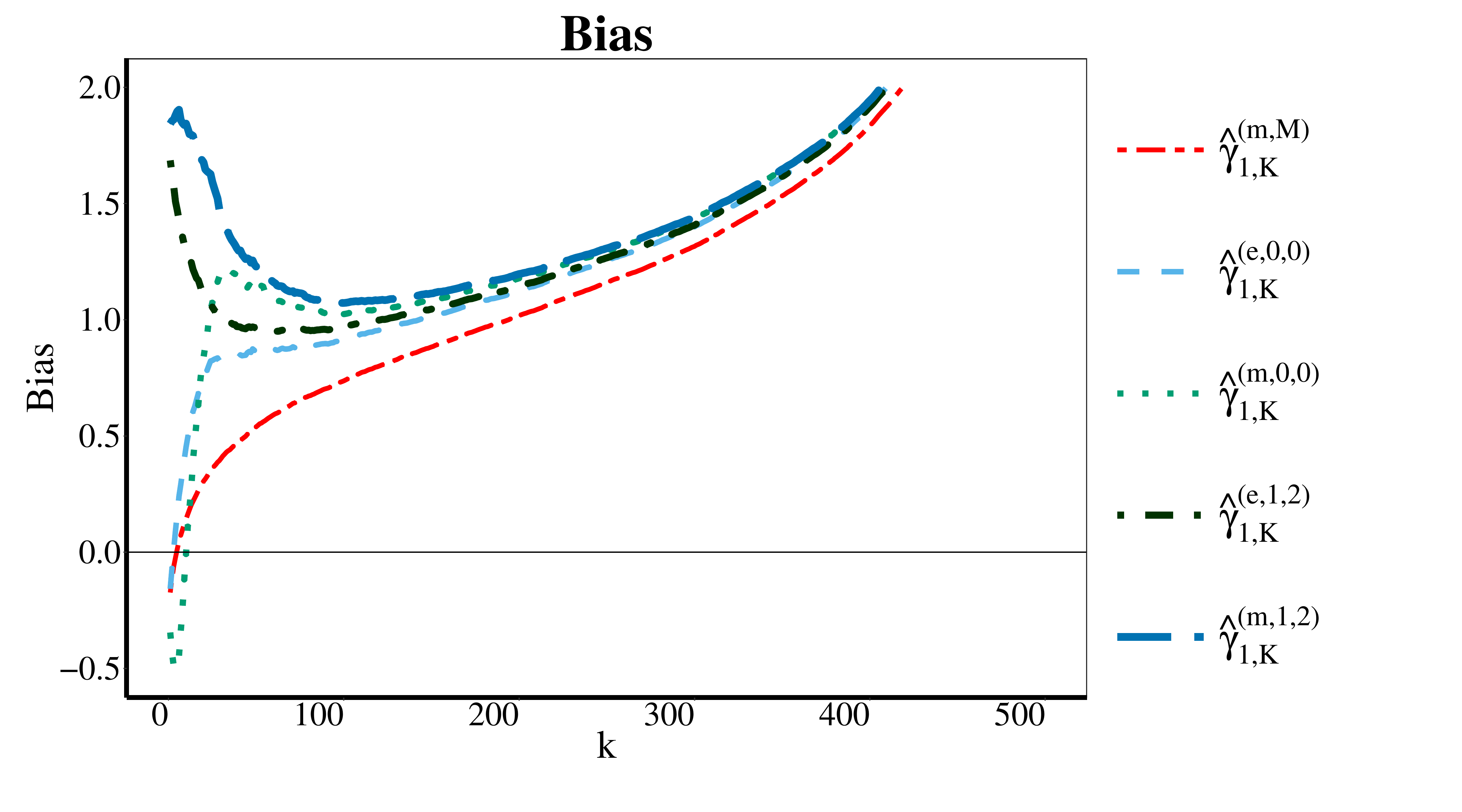}
	\end{subfigure}
	\hspace{\fill}
	\begin{subfigure}[h]{0.3\linewidth}
		\includegraphics[width=8.5cm,height=5.5cm]{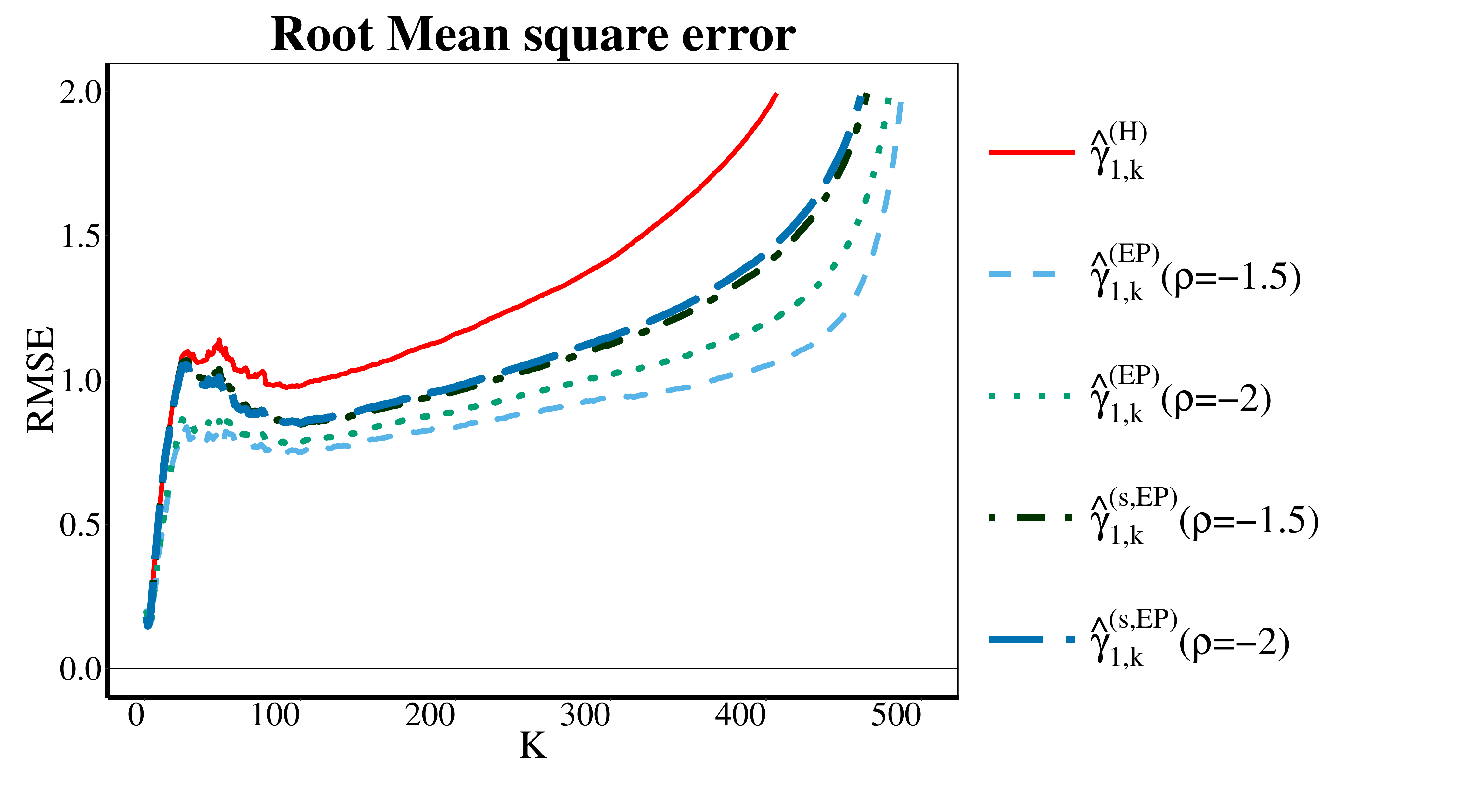}
	\end{subfigure}
		\hspace{\fill}
		\begin{subfigure}[h]{0.3\linewidth}
			\includegraphics[width=8.5cm,height=5.5cm]{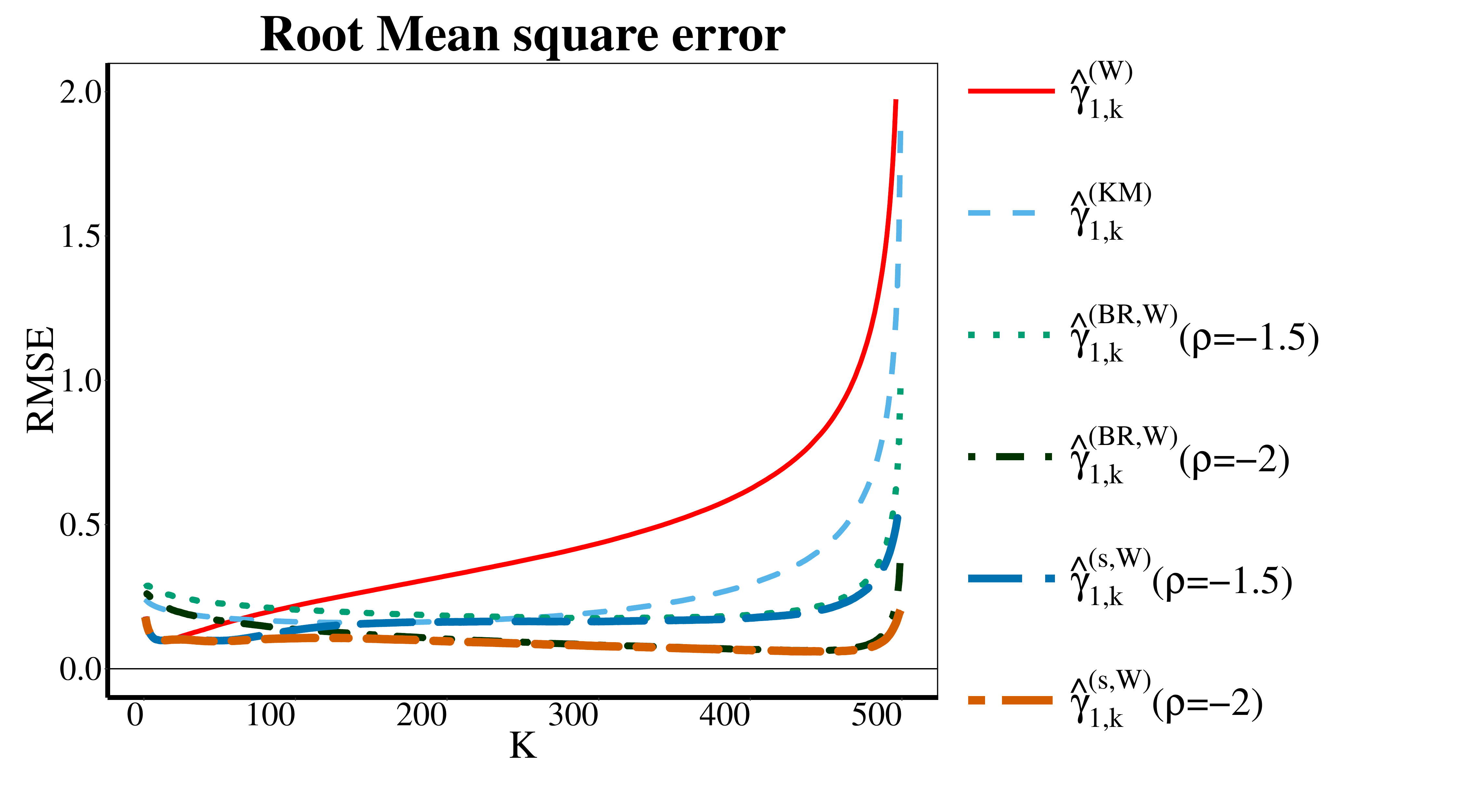}
		\end{subfigure}
		\hspace{\fill}
		\begin{subfigure}[h]{0.3\linewidth}
			\includegraphics[width=7.8cm,height=5.5cm]{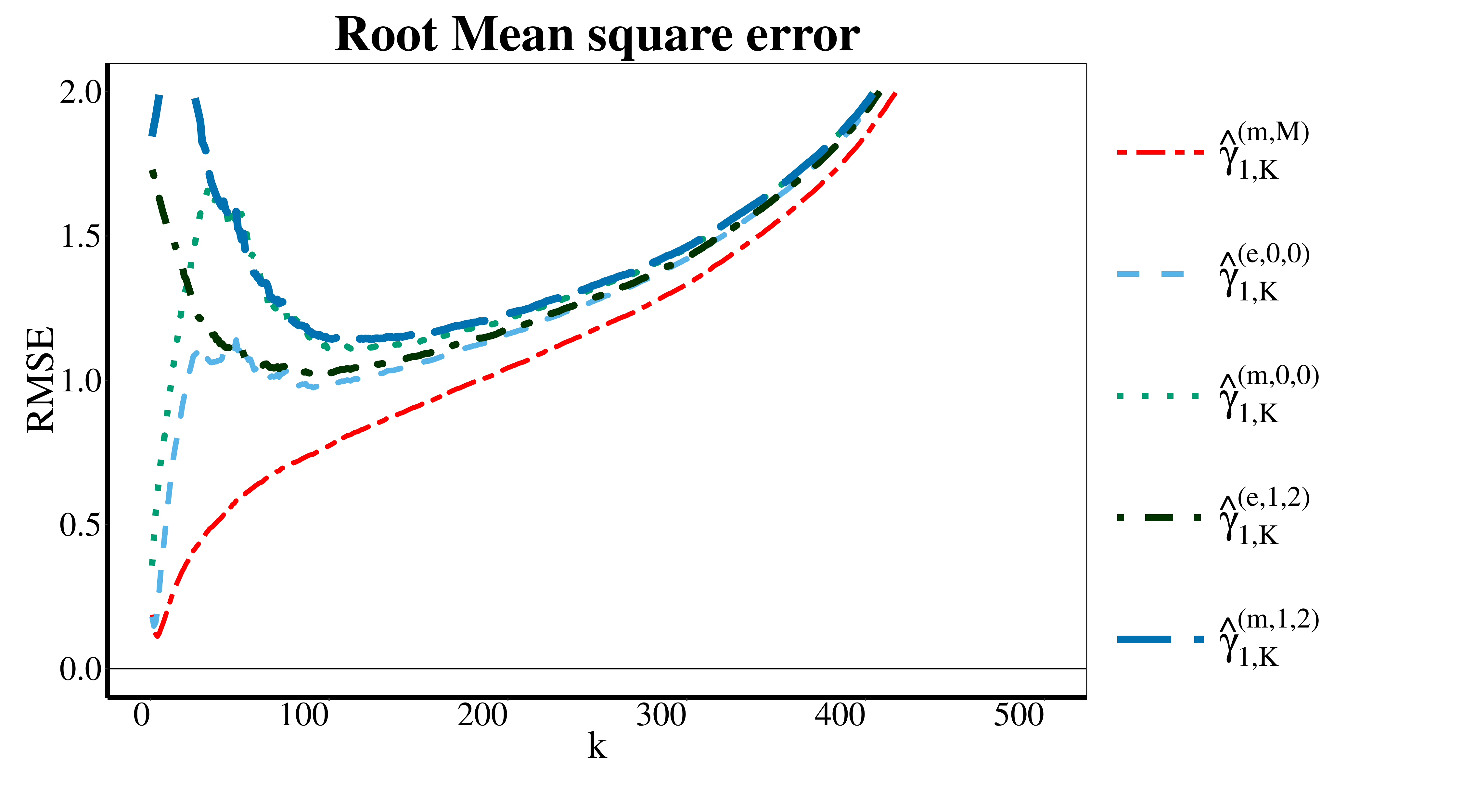}
		\end{subfigure}
		\caption{Bias and RMSE for \textbf{Burr(10,2,2)} censored by \textbf{Burr(10,5,2)}}
\end{figure}

\begin{figure}[h]
	\centering
	\begin{subfigure}[h]{0.3\linewidth}
		\includegraphics[width=8.5cm,height=5.5cm]{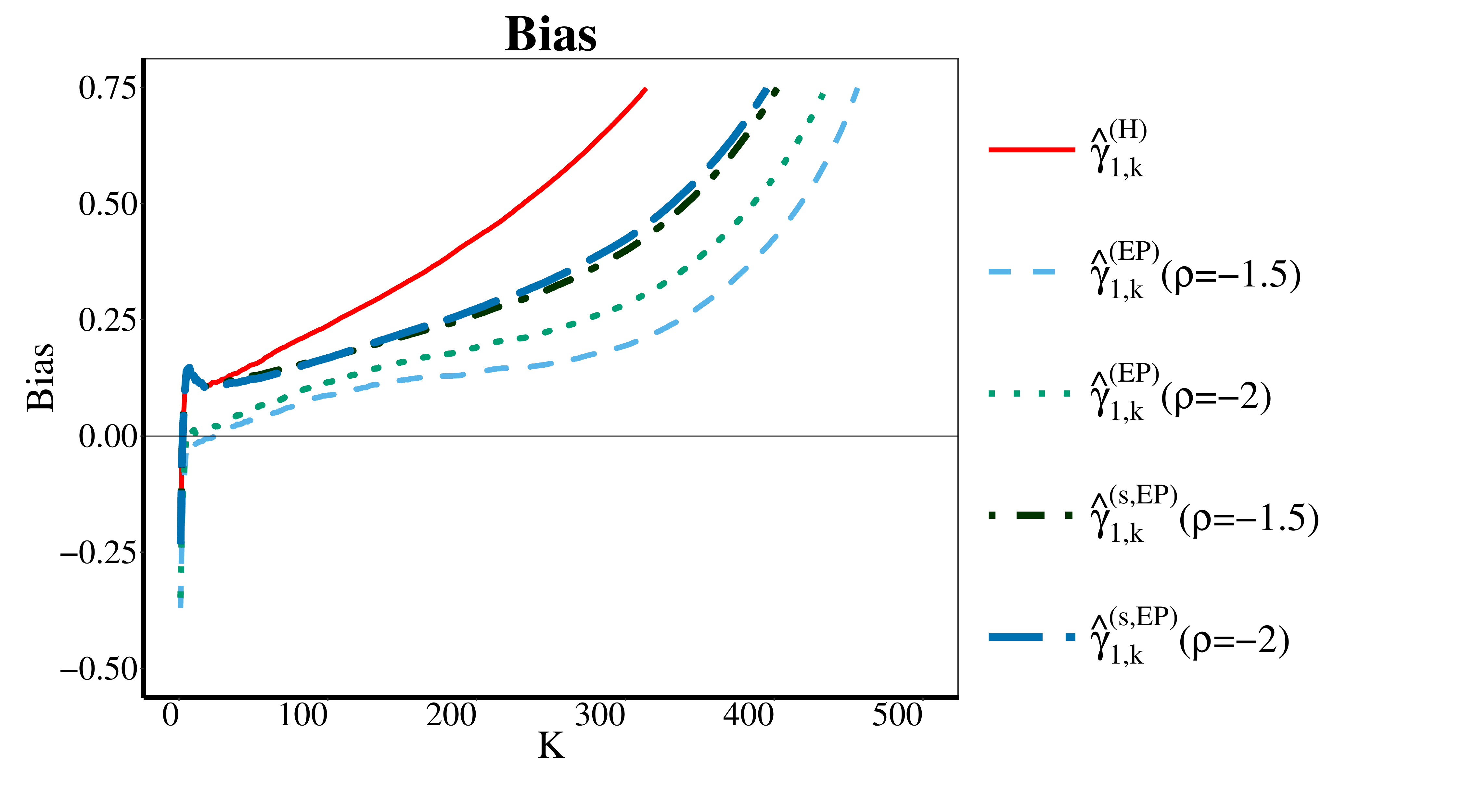}
	\end{subfigure}
	\hspace{\fill}
	\begin{subfigure}[h]{0.3\linewidth}
		\includegraphics[width=8.5cm,height=5.5cm]{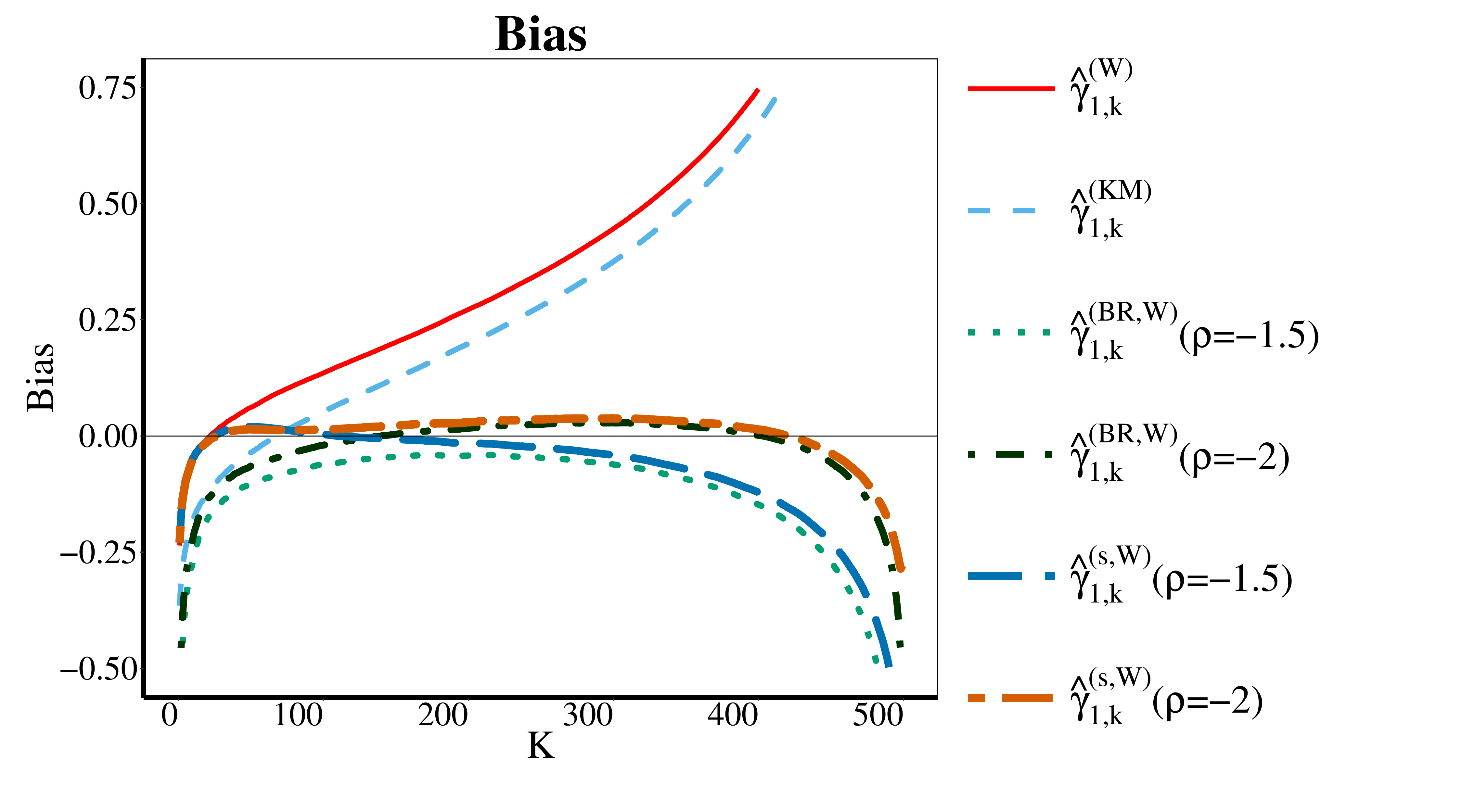}
	\end{subfigure}
	\hspace{\fill}
	\begin{subfigure}[h]{0.3\linewidth}
		\includegraphics[width=7.8cm,height=5.5cm]{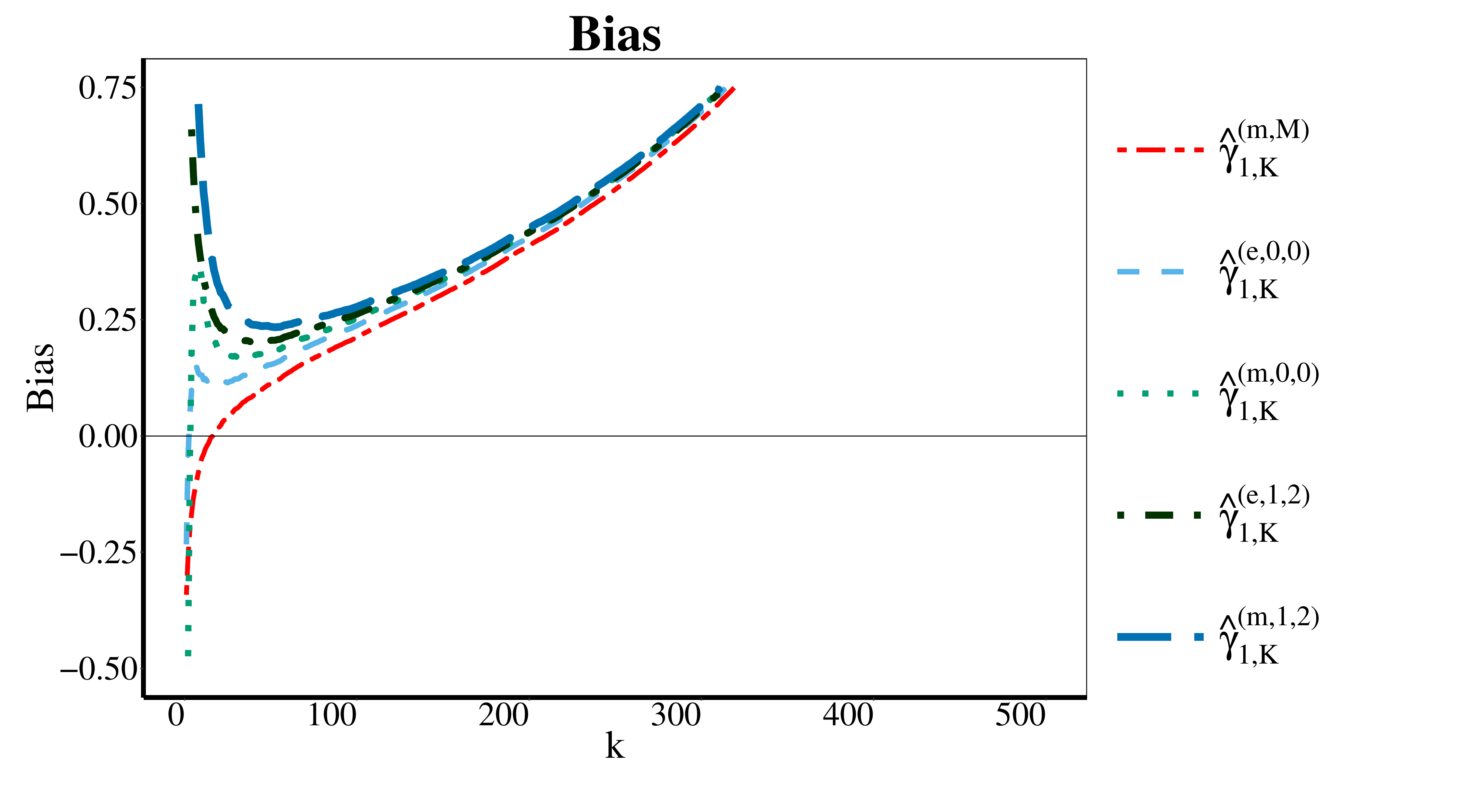}
	\end{subfigure}
	\hspace{\fill}
	\begin{subfigure}[h]{0.3\linewidth}
		\includegraphics[width=8.5cm,height=5.5cm]{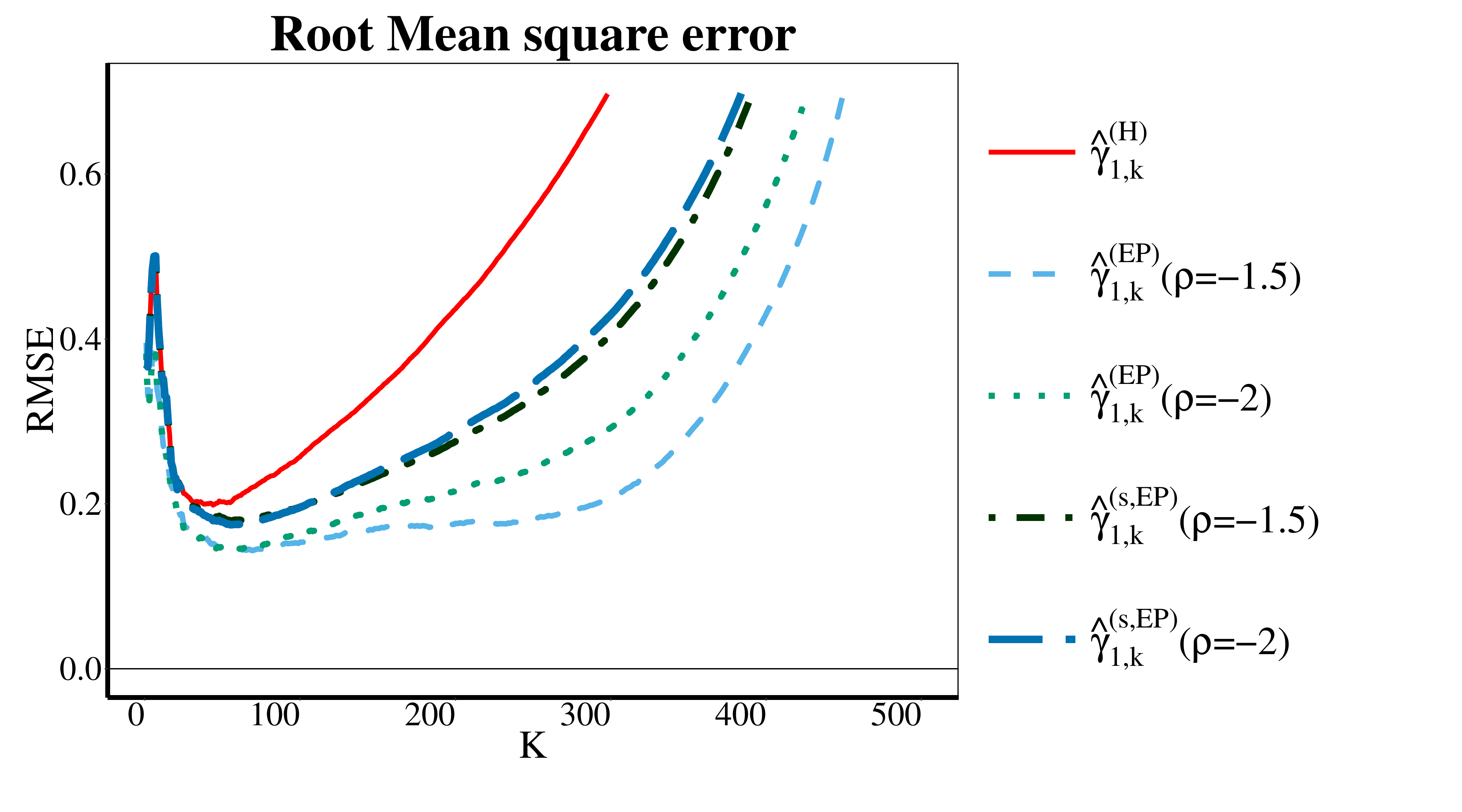}
	\end{subfigure}
	\hspace{\fill}
	\begin{subfigure}[h]{0.3\linewidth}
		\includegraphics[width=8.5cm,height=5.5cm]{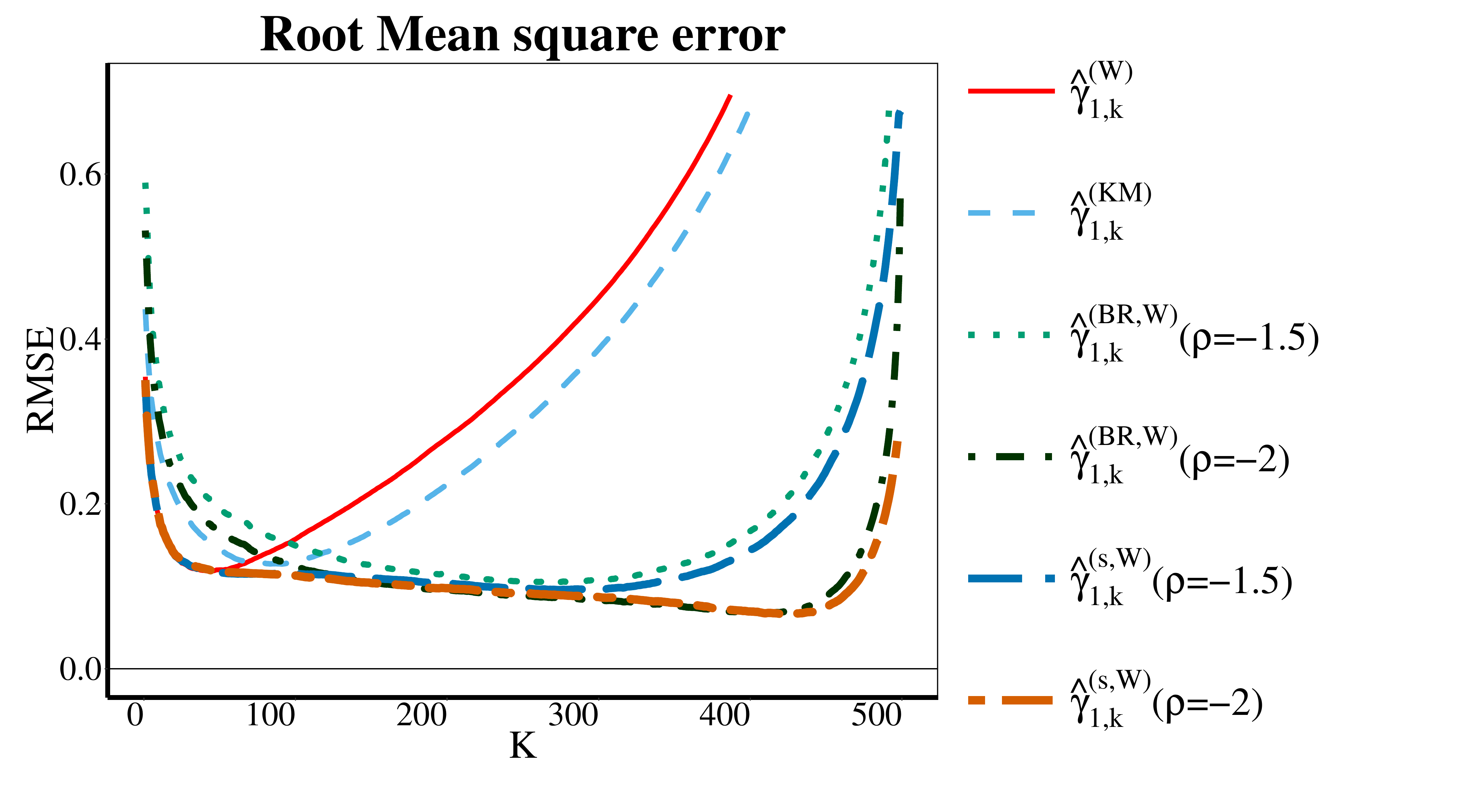}
	\end{subfigure}
	\hspace{\fill}
	\begin{subfigure}[h]{0.3\linewidth}
		\includegraphics[width=7.8cm,height=5.5cm]{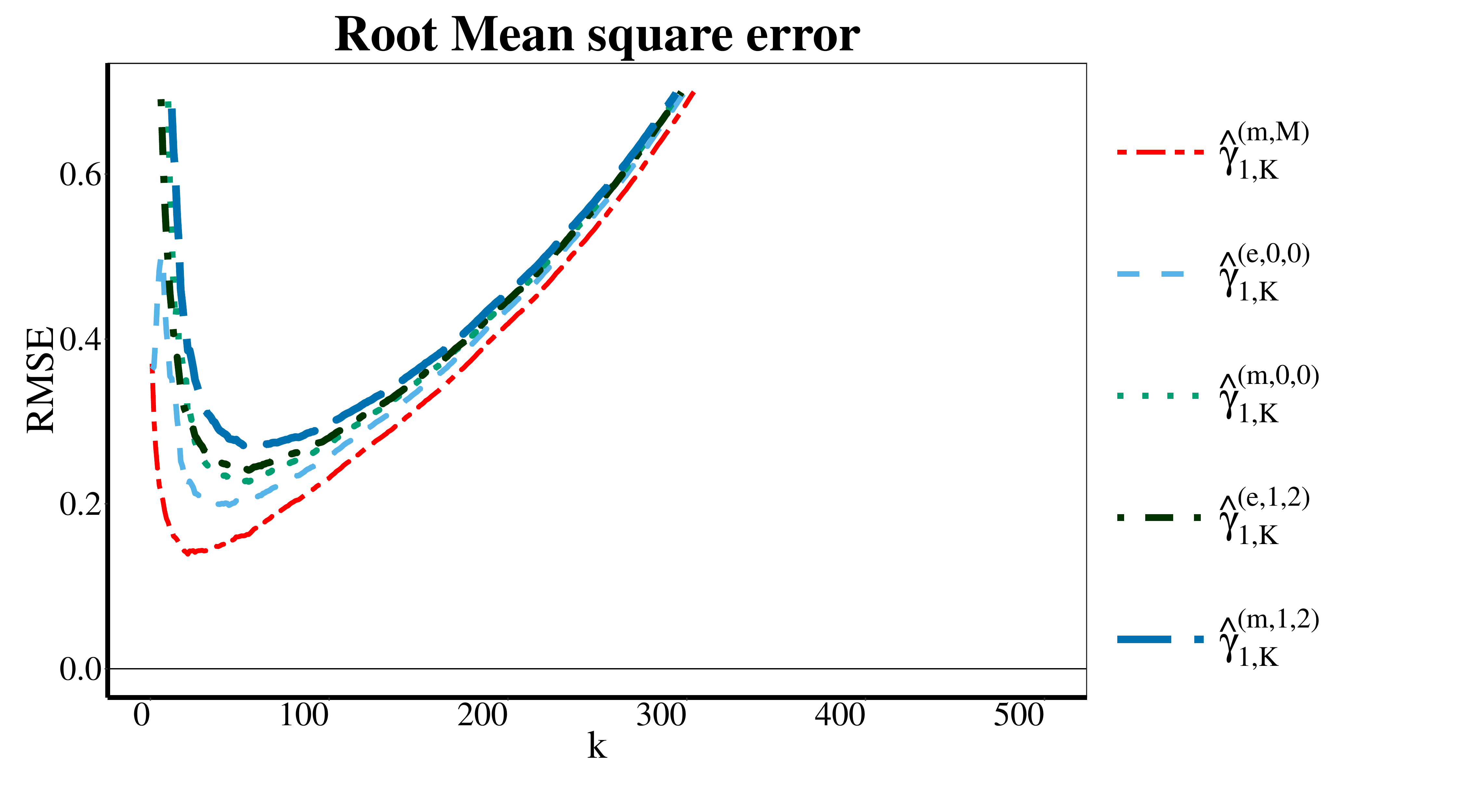}
	\end{subfigure}
	\caption{Bias and RMSE for \textbf{Burr(10,2,1)} censored by \textbf{Burr(10,2,1)}}
\end{figure}

\begin{figure}[h]
	\centering
	\begin{subfigure}[h]{0.3\linewidth}
		\includegraphics[width=8.5cm,height=5.5cm]{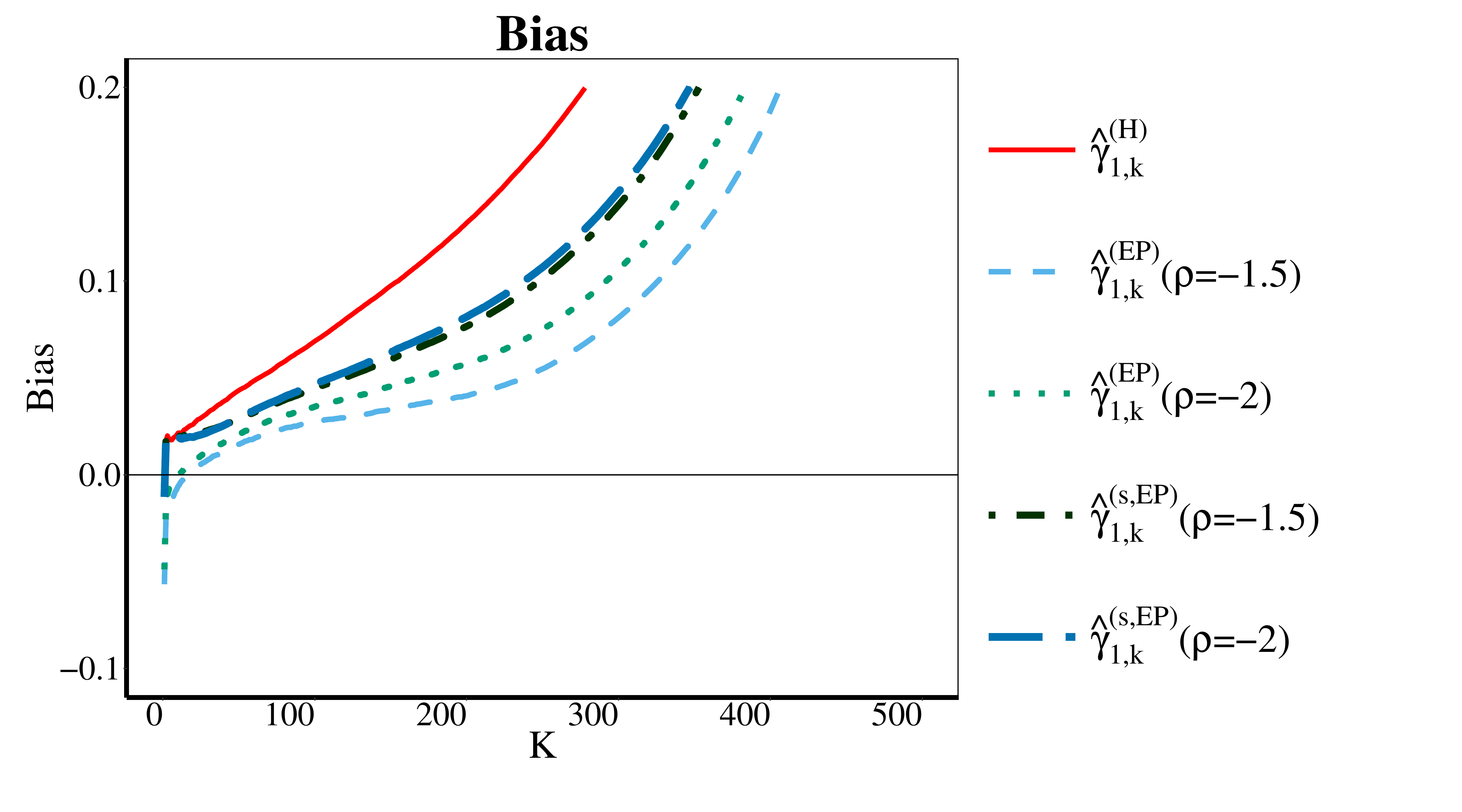}
	\end{subfigure}
	\hspace{\fill}
	\begin{subfigure}[h]{0.3\linewidth}
		\includegraphics[width=8.5cm,height=5.5cm]{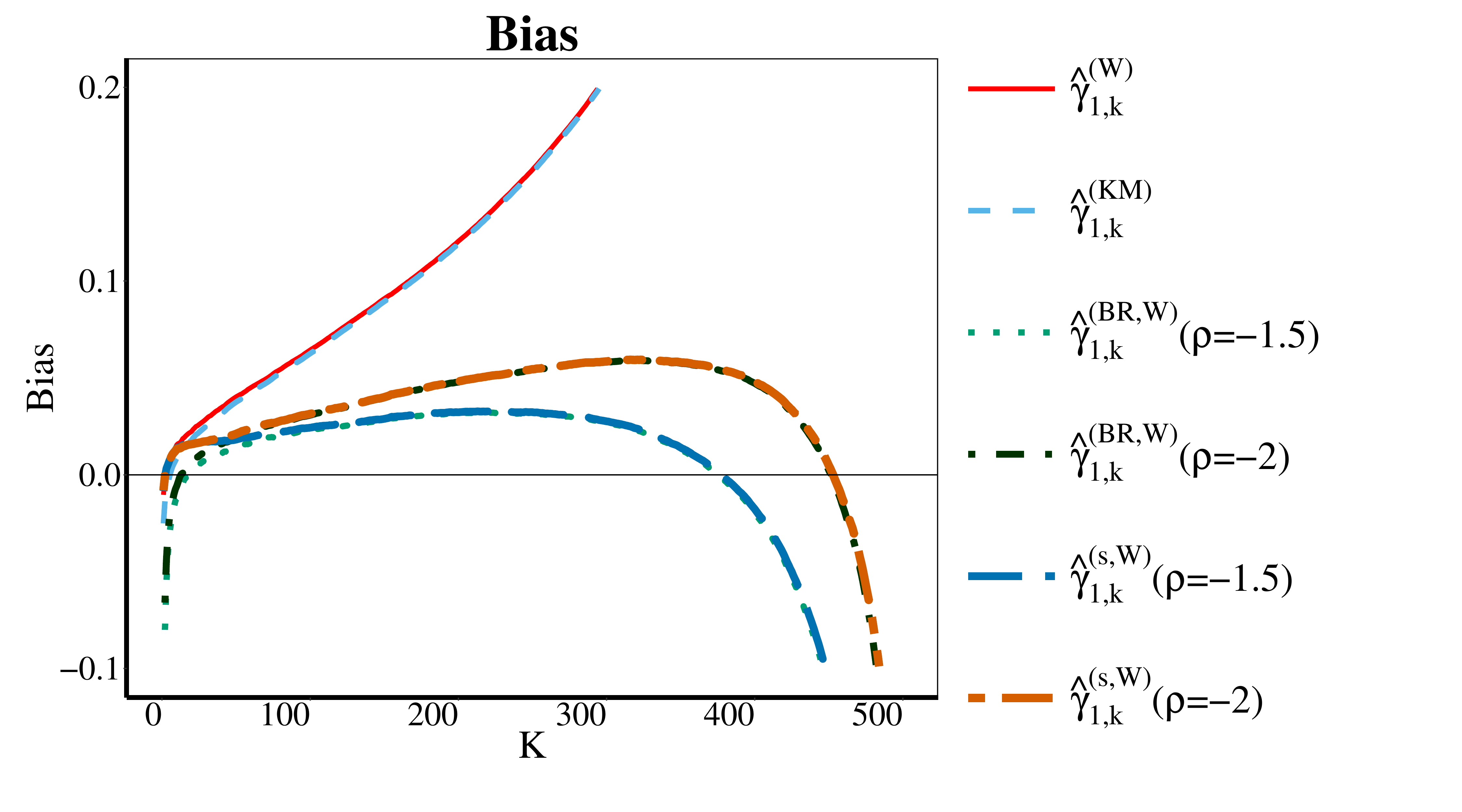}
	\end{subfigure}
	\hspace{\fill}
	\begin{subfigure}[h]{0.3\linewidth}
		\includegraphics[width=7.8cm,height=5.5cm]{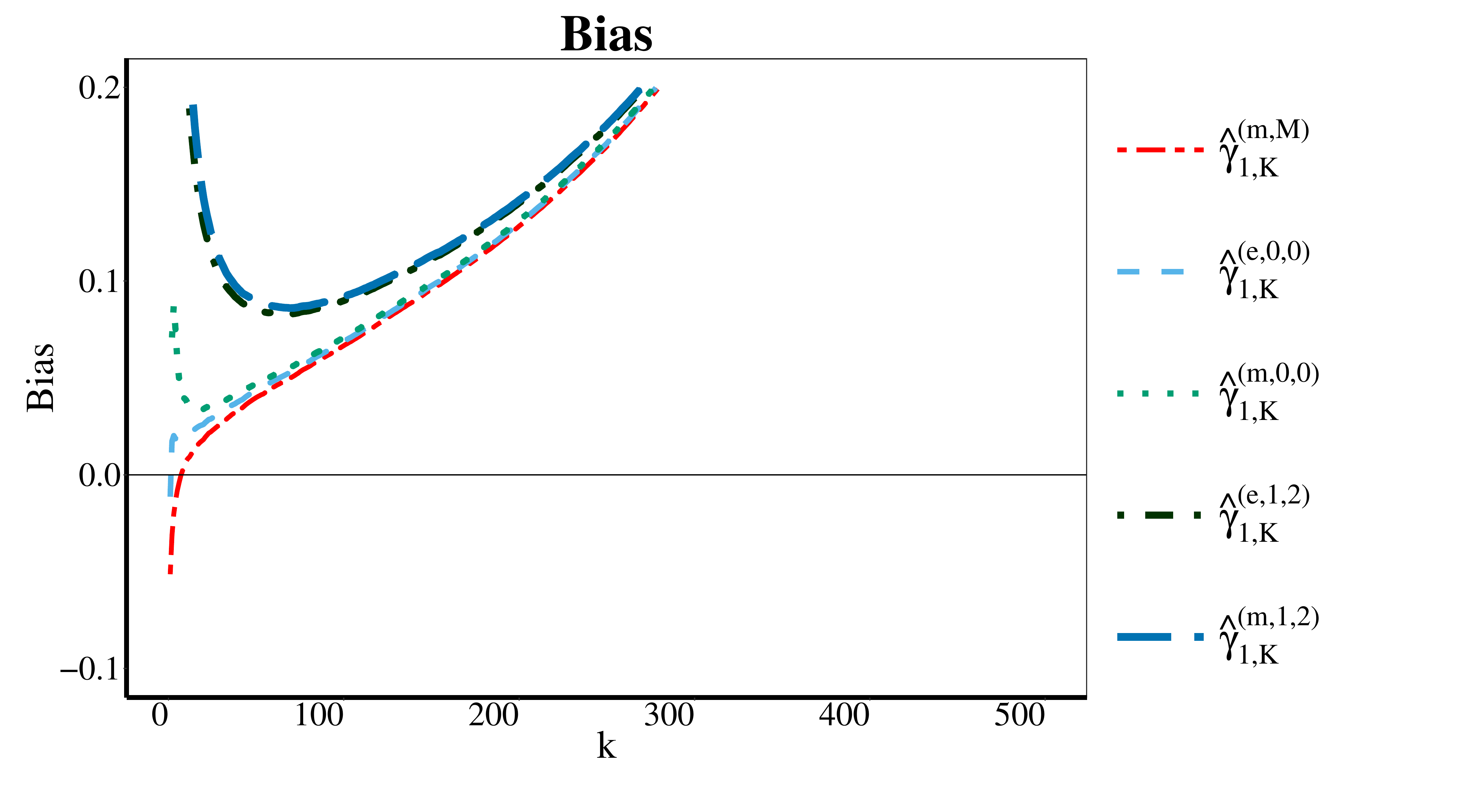}
	\end{subfigure}
	\hspace{\fill}
	\begin{subfigure}[h]{0.3\linewidth}
		\includegraphics[width=8.5cm,height=5.5cm]{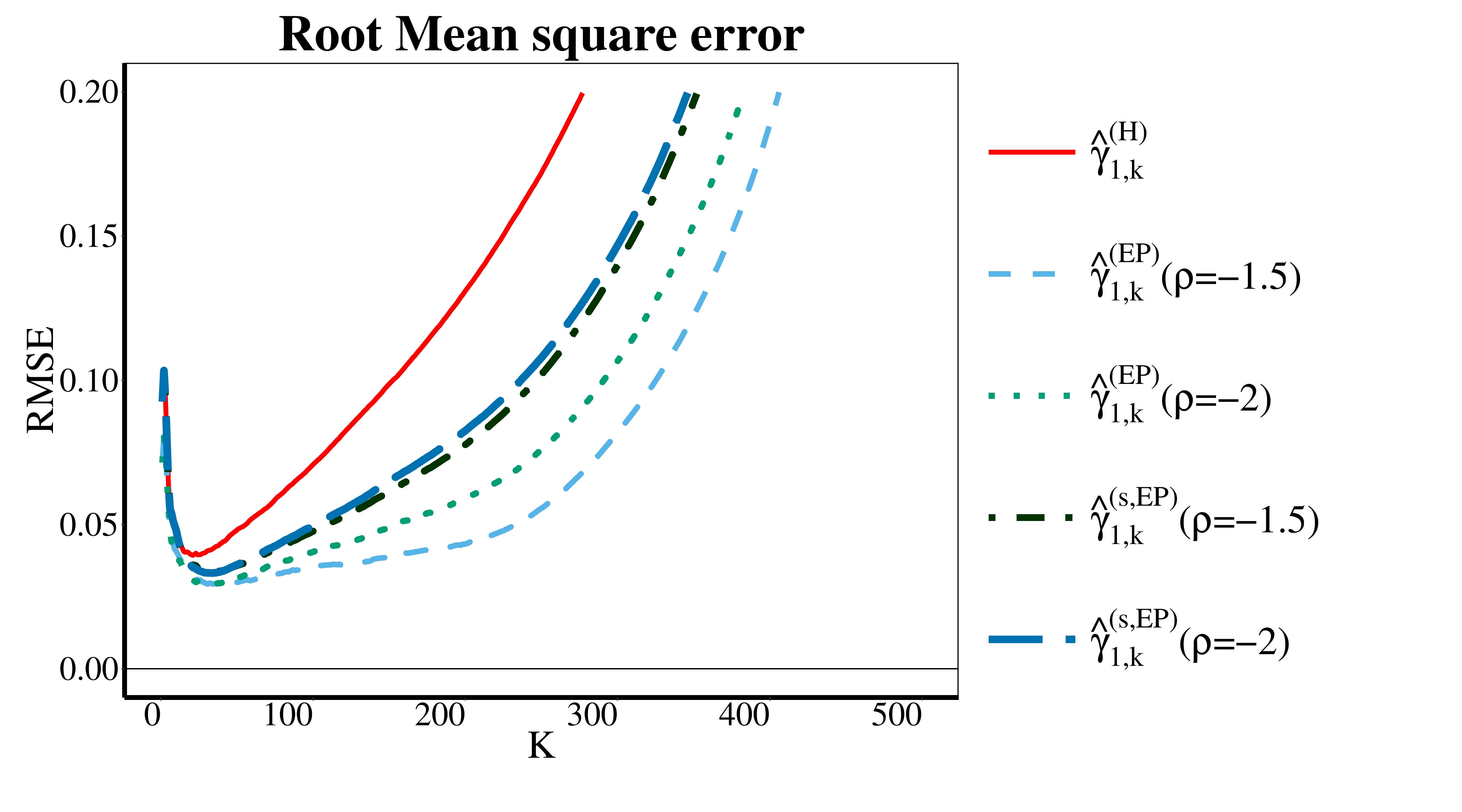}
	\end{subfigure}
	\hspace{\fill}
	\begin{subfigure}[h]{0.3\linewidth}
		\includegraphics[width=8.5cm,height=5.5cm]{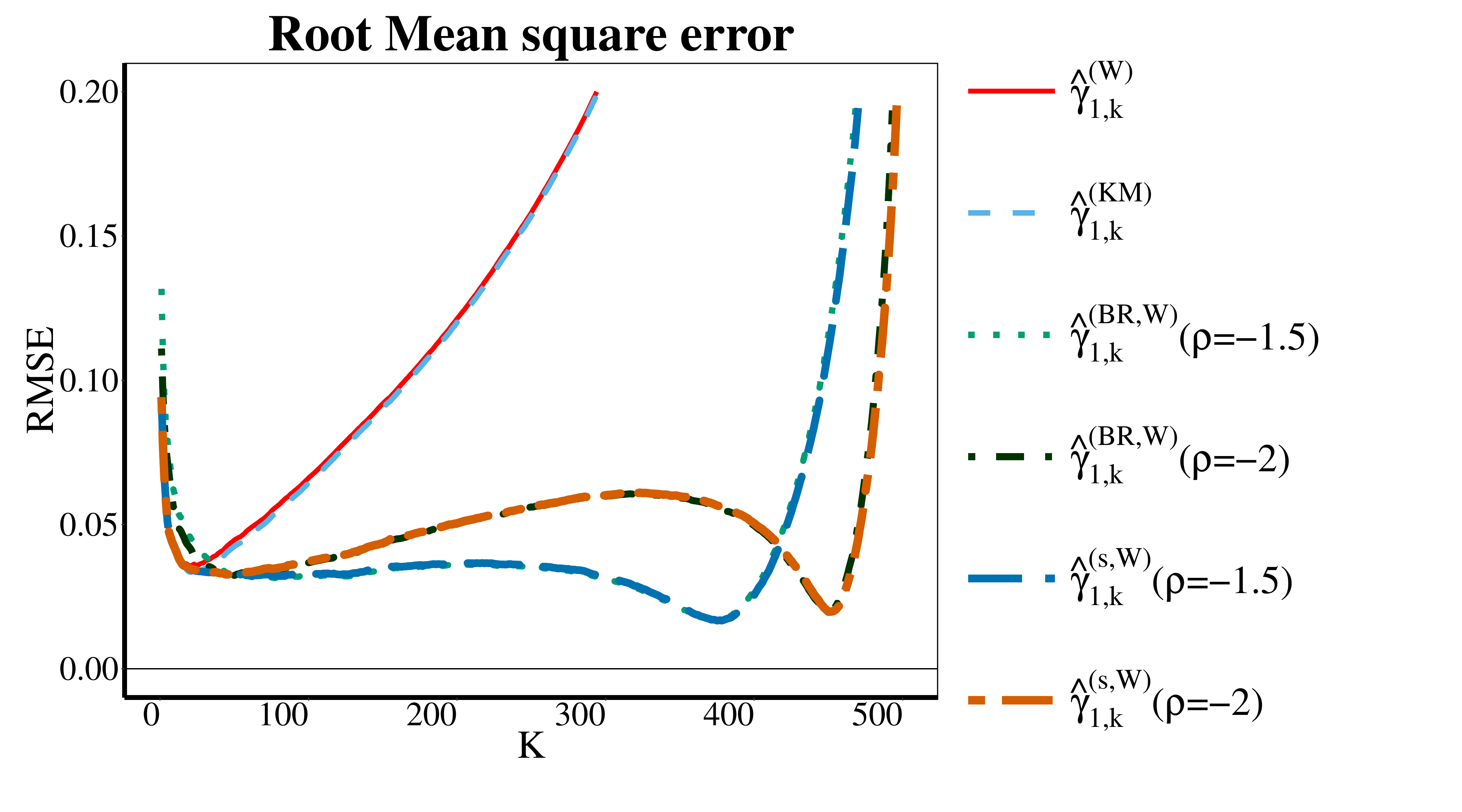}
	\end{subfigure}
	\hspace{\fill}
	\begin{subfigure}[h]{0.3\linewidth}
		\includegraphics[width=7.8cm,height=5.5cm]{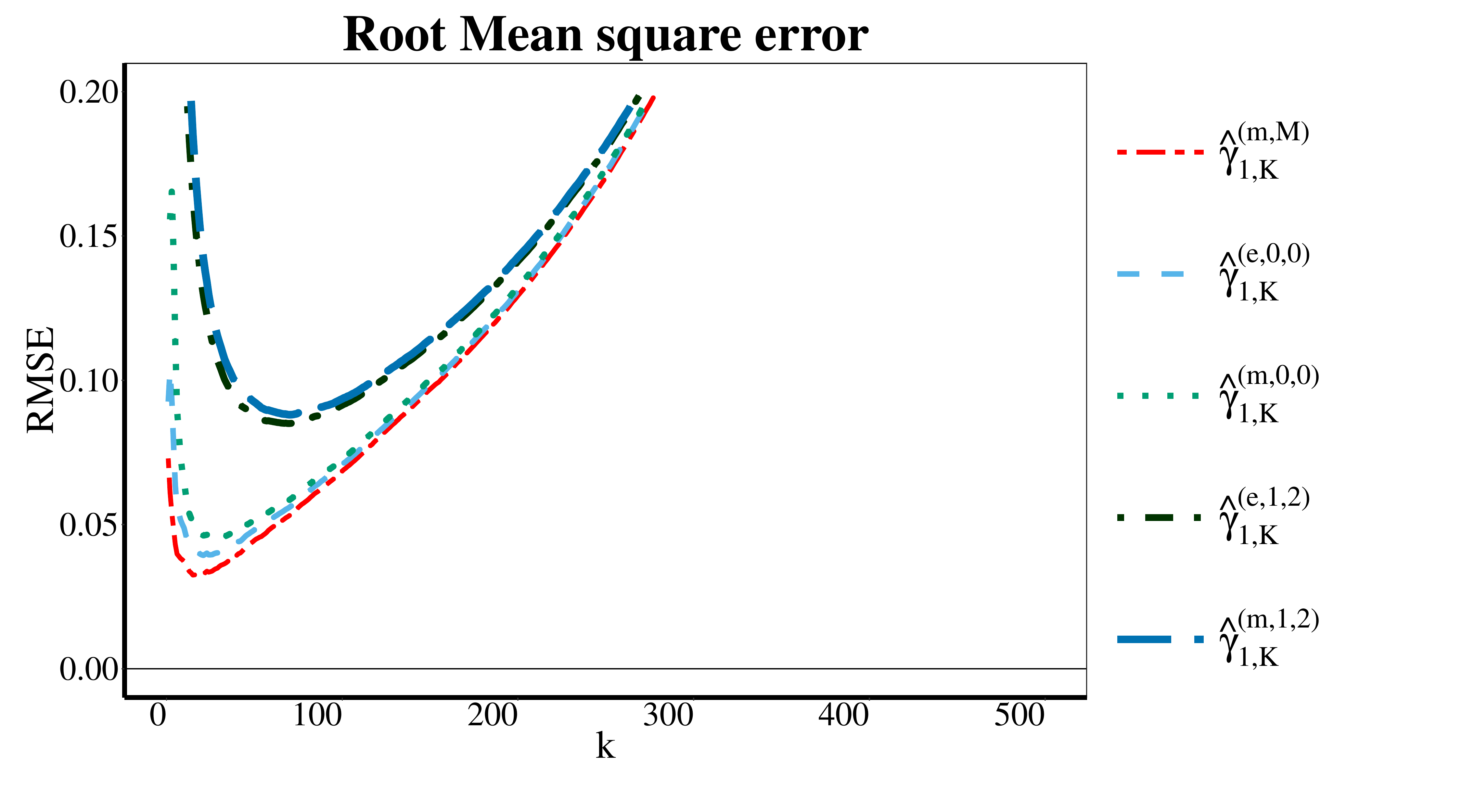}
	\end{subfigure}
	\caption{Bias and RMSE for \textbf{Burr(10,5,2)} censored by \textbf{Burr(10,2,2)}}
\end{figure}


\begin{figure}[h]
	\centering
	\begin{subfigure}[h]{0.3\linewidth}
		\includegraphics[width=8.5cm,height=5.5cm]{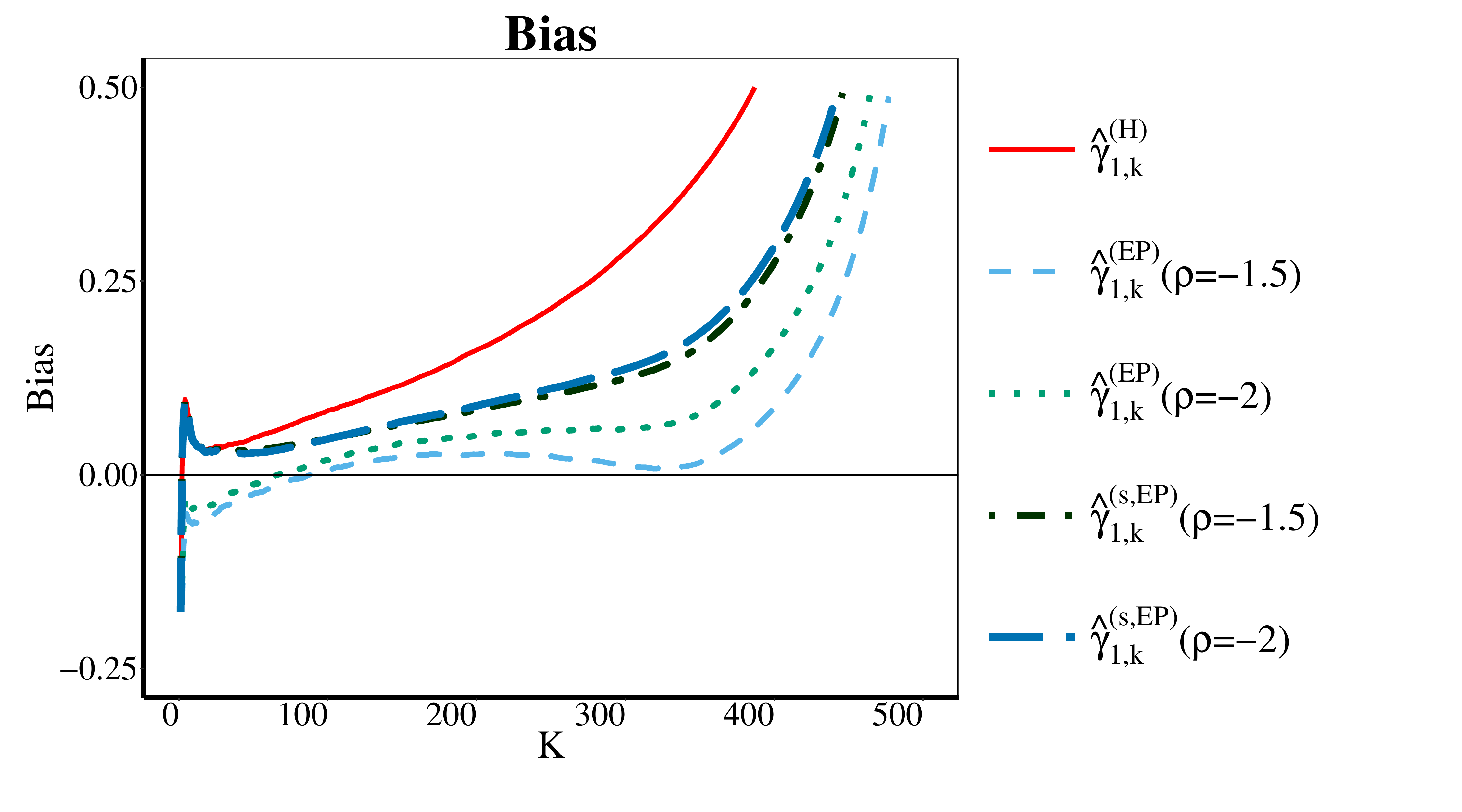}
	\end{subfigure}
	\hspace{\fill}
	\begin{subfigure}[h]{0.3\linewidth}
		\includegraphics[width=8.5cm,height=5.5cm]{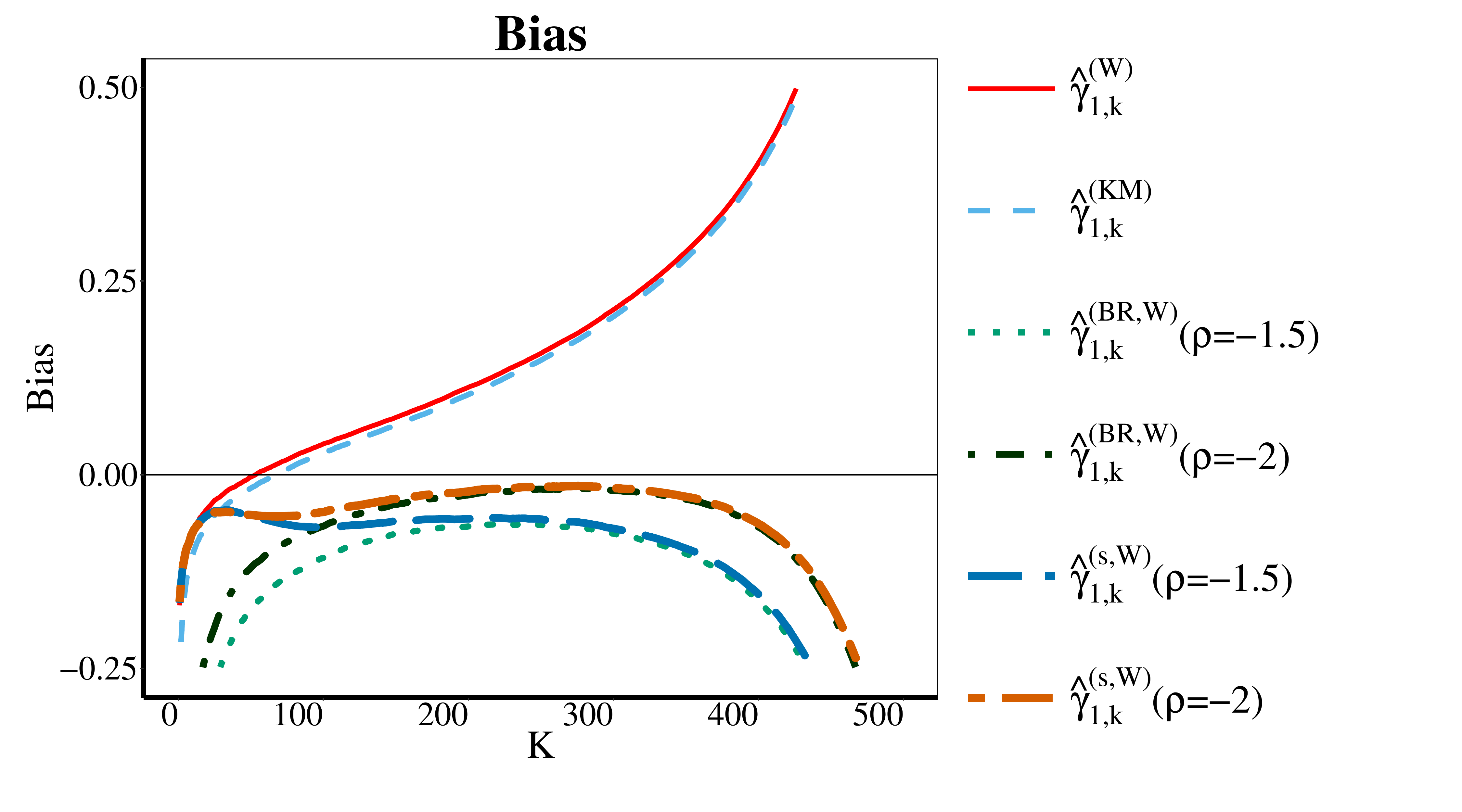}
	\end{subfigure}
	\hspace{\fill}
	\begin{subfigure}[h]{0.3\linewidth}
		\includegraphics[width=7.8cm,height=5.5cm]{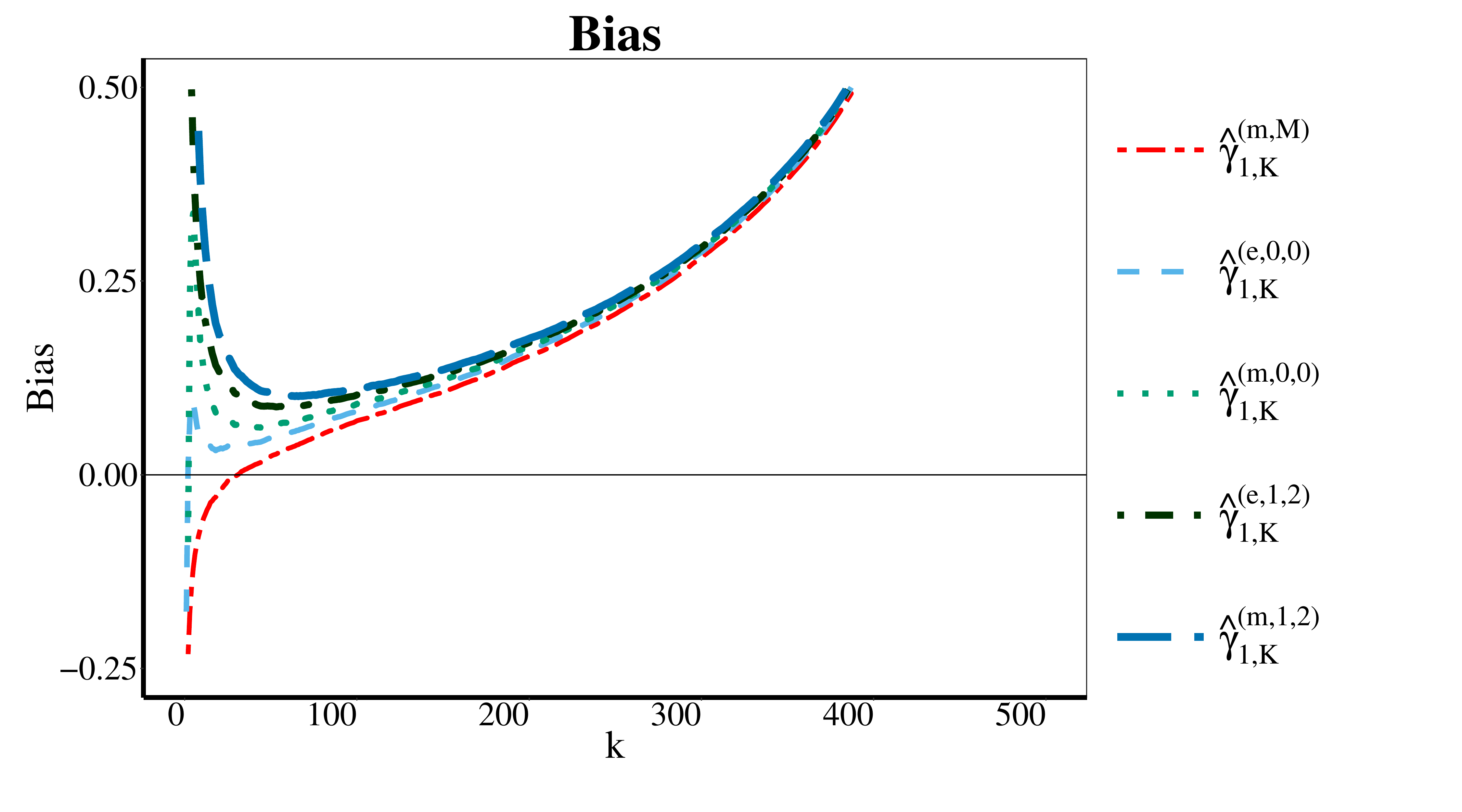}
	\end{subfigure}
	\hspace{\fill}
	\begin{subfigure}[h]{0.3\linewidth}
		\includegraphics[width=8.5cm,height=5.5cm]{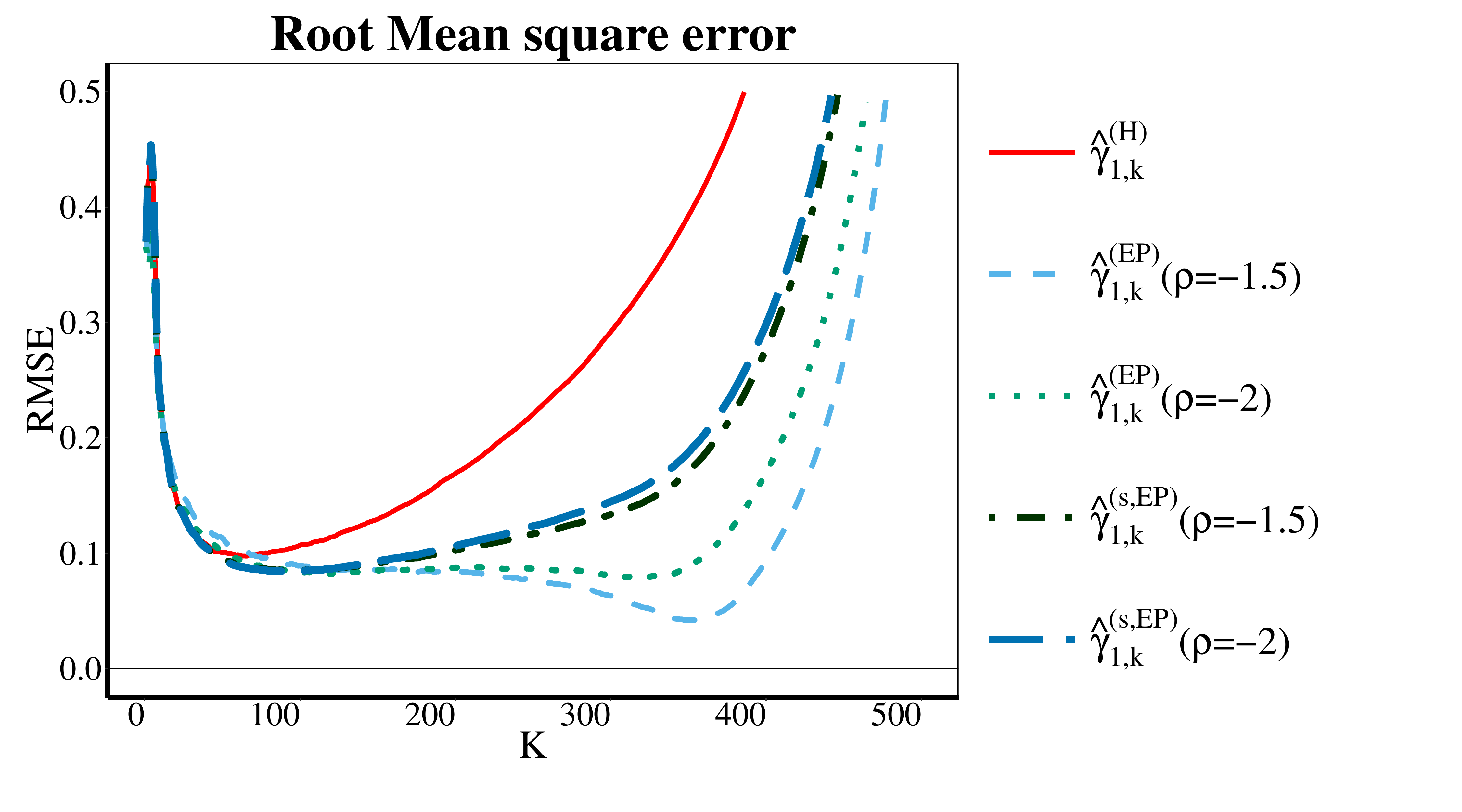}
	\end{subfigure}
	\hspace{\fill}
	\begin{subfigure}[h]{0.3\linewidth}
		\includegraphics[width=8.5cm,height=5.5cm]{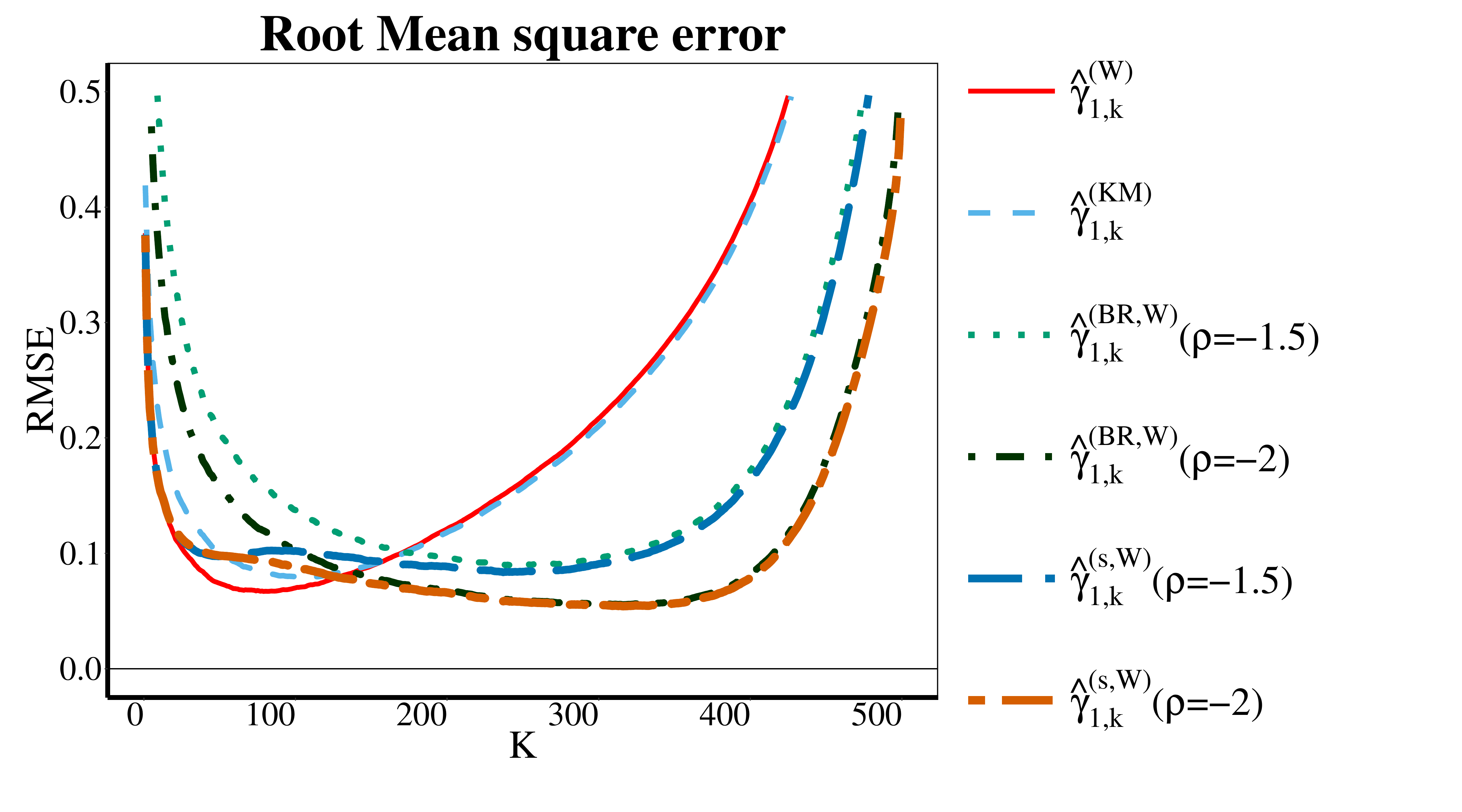}
	\end{subfigure}
	\hspace{\fill}
	\begin{subfigure}[h]{0.3\linewidth}
		\includegraphics[width=8.5cm,height=5.5cm]{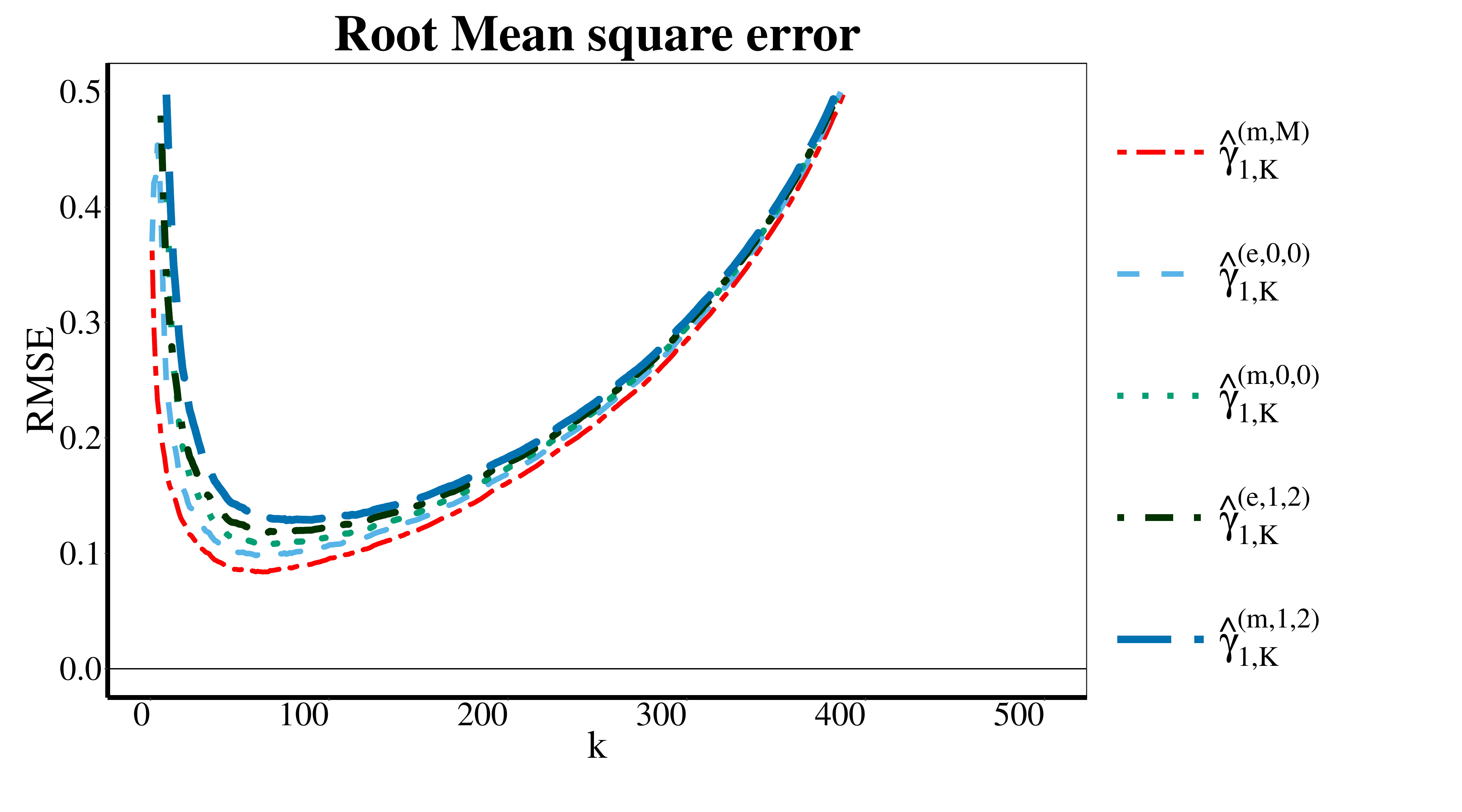}
	\end{subfigure}
	\caption{Bias and RMSE for \textbf{Fr\'echet(2)} censored by \textbf{Fr\'echet(1)}}
\end{figure}


\begin{figure}[h]
	\centering
	\begin{subfigure}[h]{0.3\linewidth}
		\includegraphics[width=8.5cm,height=5.5cm]{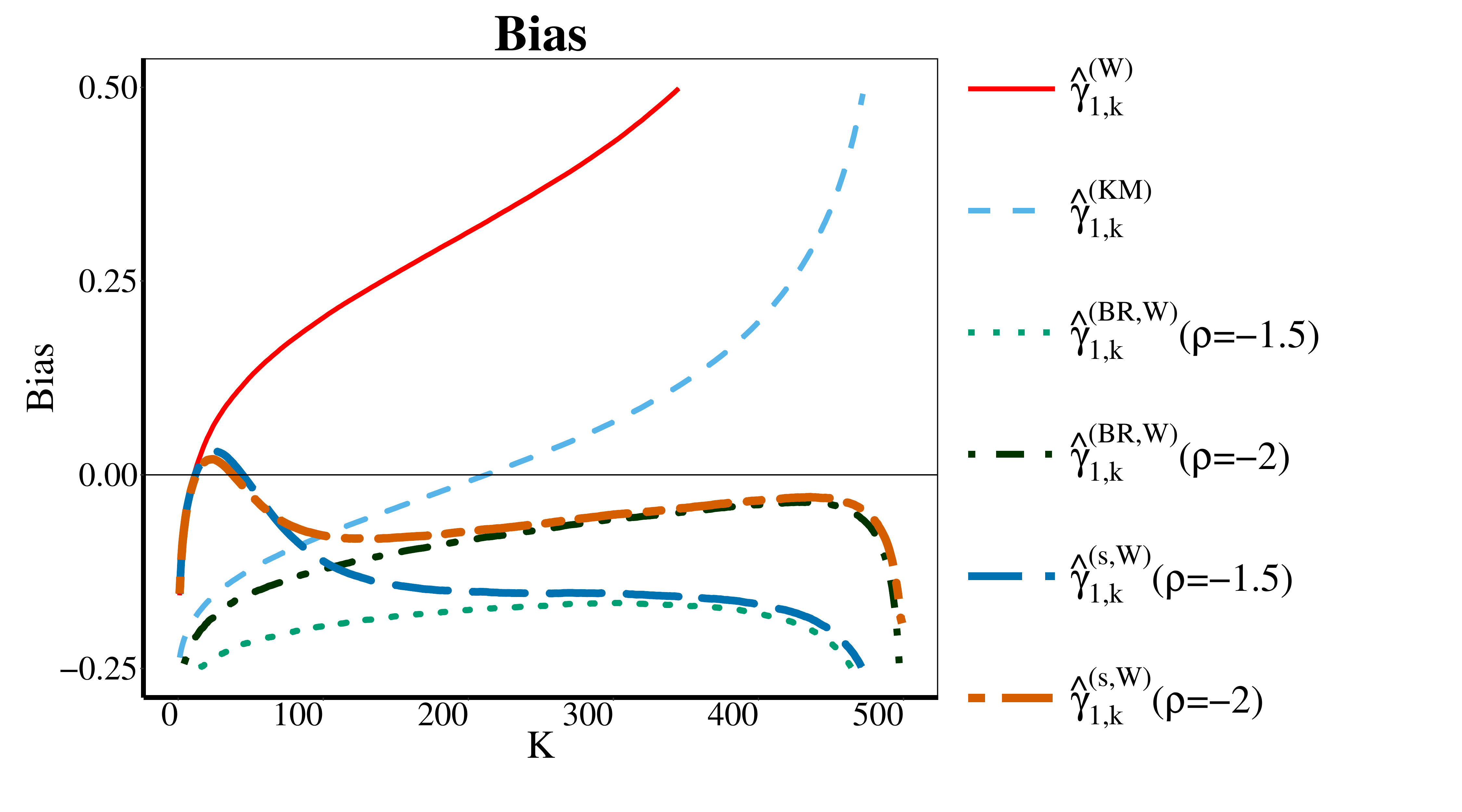}
	\end{subfigure}
	\hspace{\fill}
	\begin{subfigure}[h]{0.3\linewidth}
		\includegraphics[width=8.5cm,height=5.5cm]{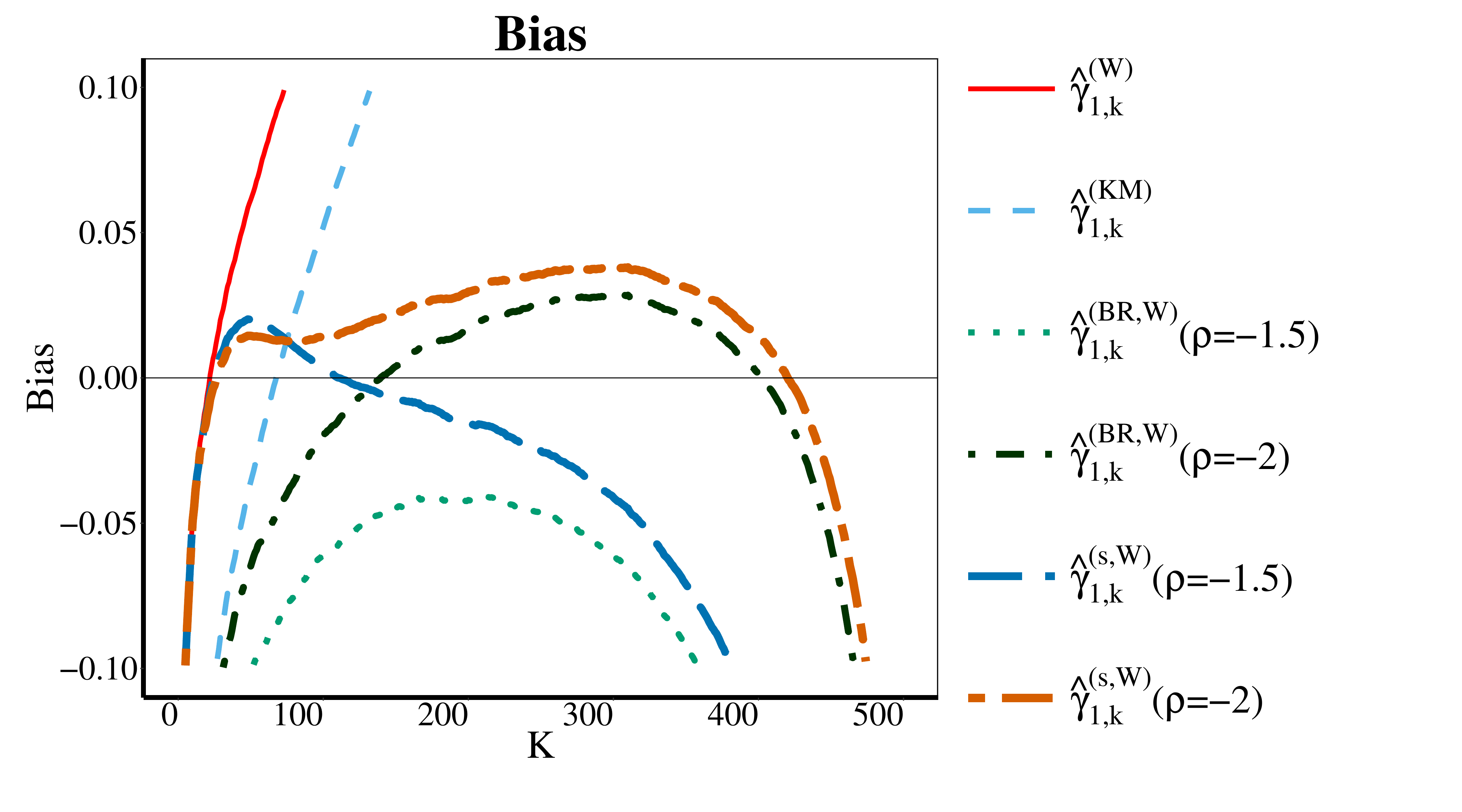}
	\end{subfigure}
	\hspace{\fill}
	\begin{subfigure}[h]{0.3\linewidth}
		\includegraphics[width=8.5cm,height=5.5cm]{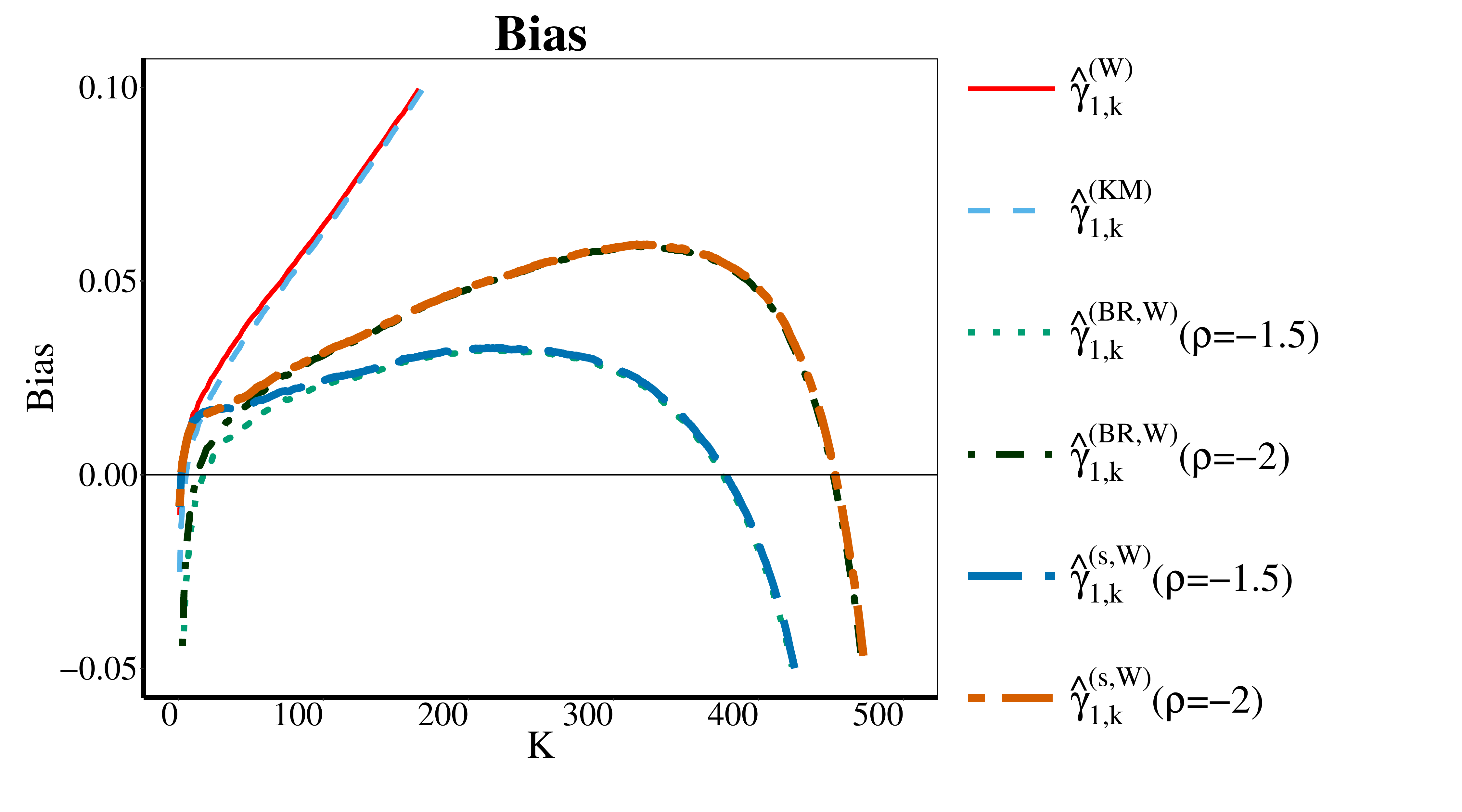}
	\end{subfigure}
	\hspace{\fill}
	\begin{subfigure}[h]{0.3\linewidth}
		\includegraphics[width=8.5cm,height=5.5cm]{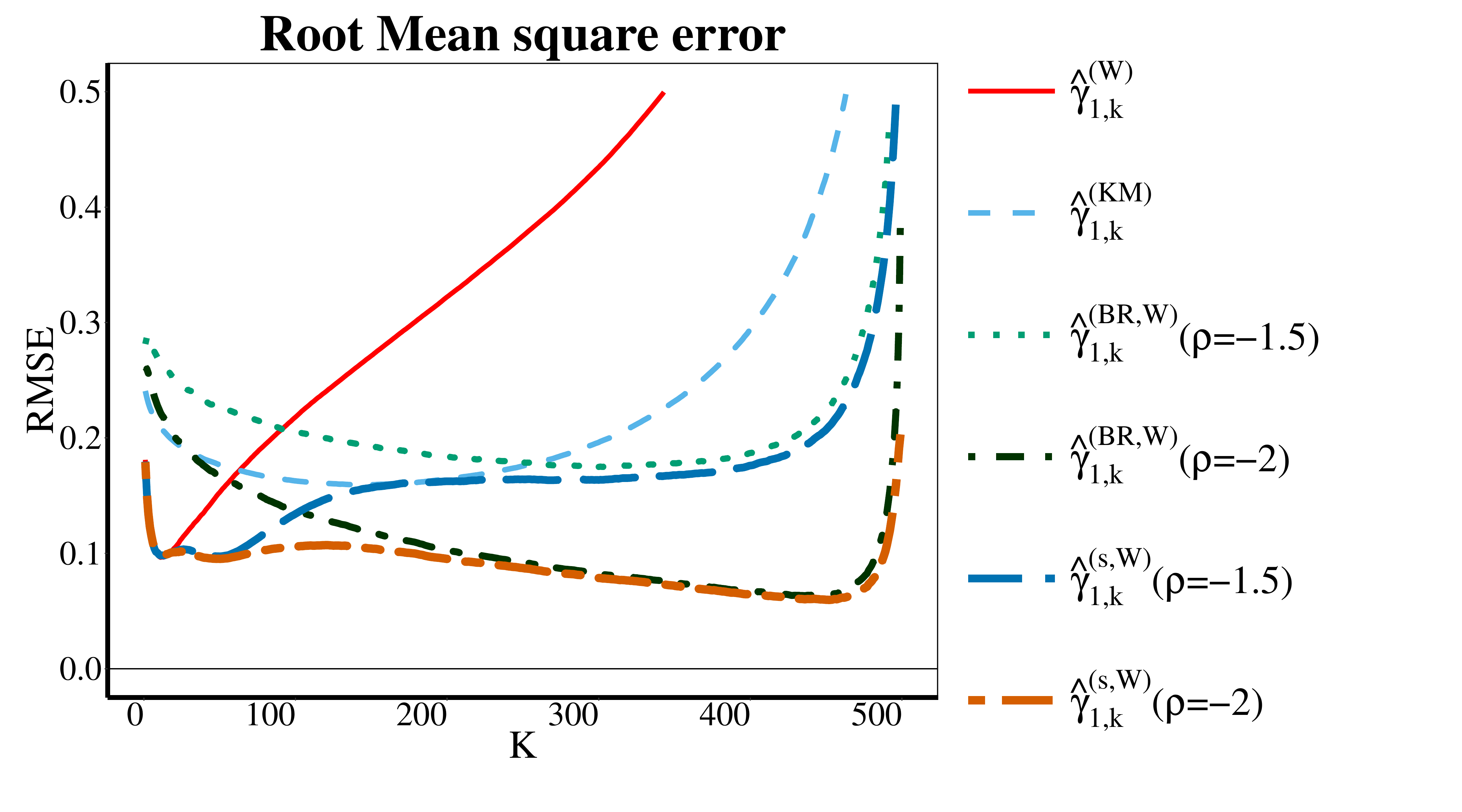}
	\end{subfigure}
	\hspace{\fill}
	\begin{subfigure}[h]{0.3\linewidth}
		\includegraphics[width=8.5cm,height=5.5cm]{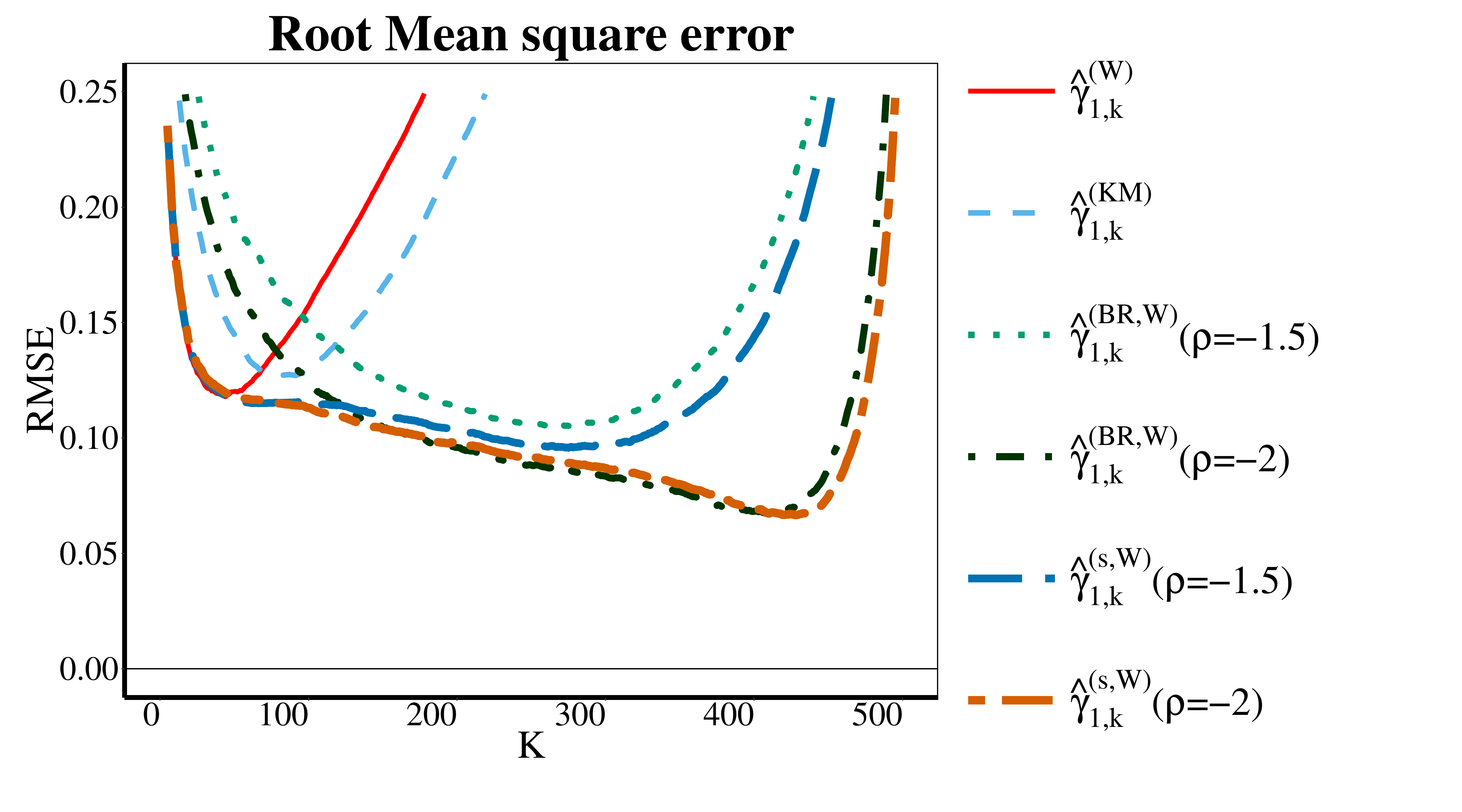}
	\end{subfigure}
	\hspace{\fill}
	\begin{subfigure}[h]{0.3\linewidth}
		\includegraphics[width=8.5cm,height=5.5cm]{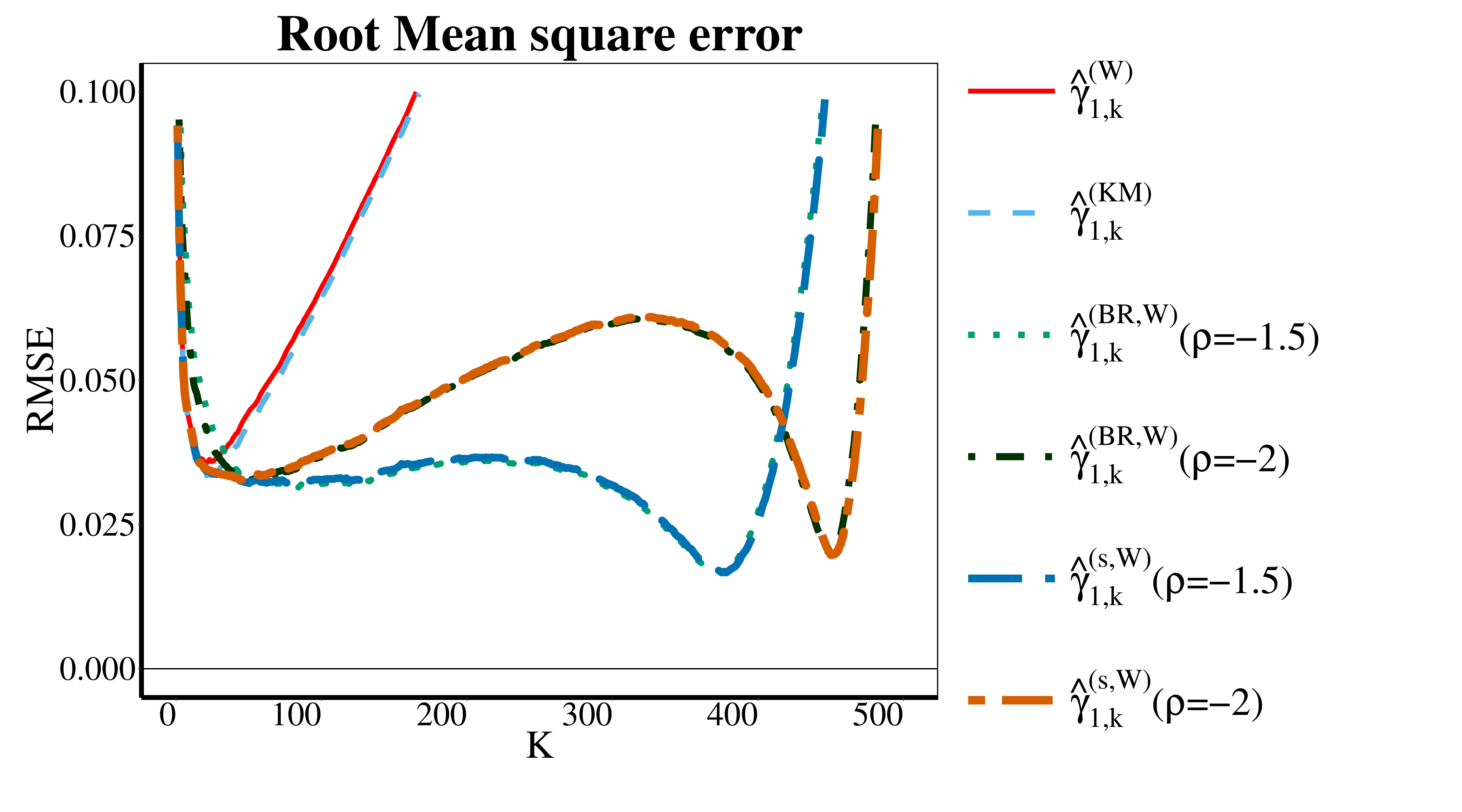}
	\end{subfigure}
	\caption{Bias (top) and RMSE (bottom) for $\hat{\gamma}_{1,k}^{(W)}$, $\hat{\gamma}_{1,k}^{(BR,W)}(-1.5)$, $\hat{\gamma}_{1,k}^{(BR,W)}(-2)$,  $\hat{\gamma}_{1,k}^{(s,W)}(-1.5)$, $\hat{\gamma}_{1,k}^{(s,W)}(-2)$ in case of \textbf{Burr(10,2,2)} censored by \textbf{Burr(10,5,2)} (left); \textbf{Burr(10,2,1)} censored by \textbf{Burr(10,2,1)} (middle); and \textbf{Burr(10,5,2)} censored by \textbf{Burr(10,2,2)} (right) }
\end{figure}

\begin{figure}[h]
		\centering
		\includegraphics[width=0.6\textwidth]{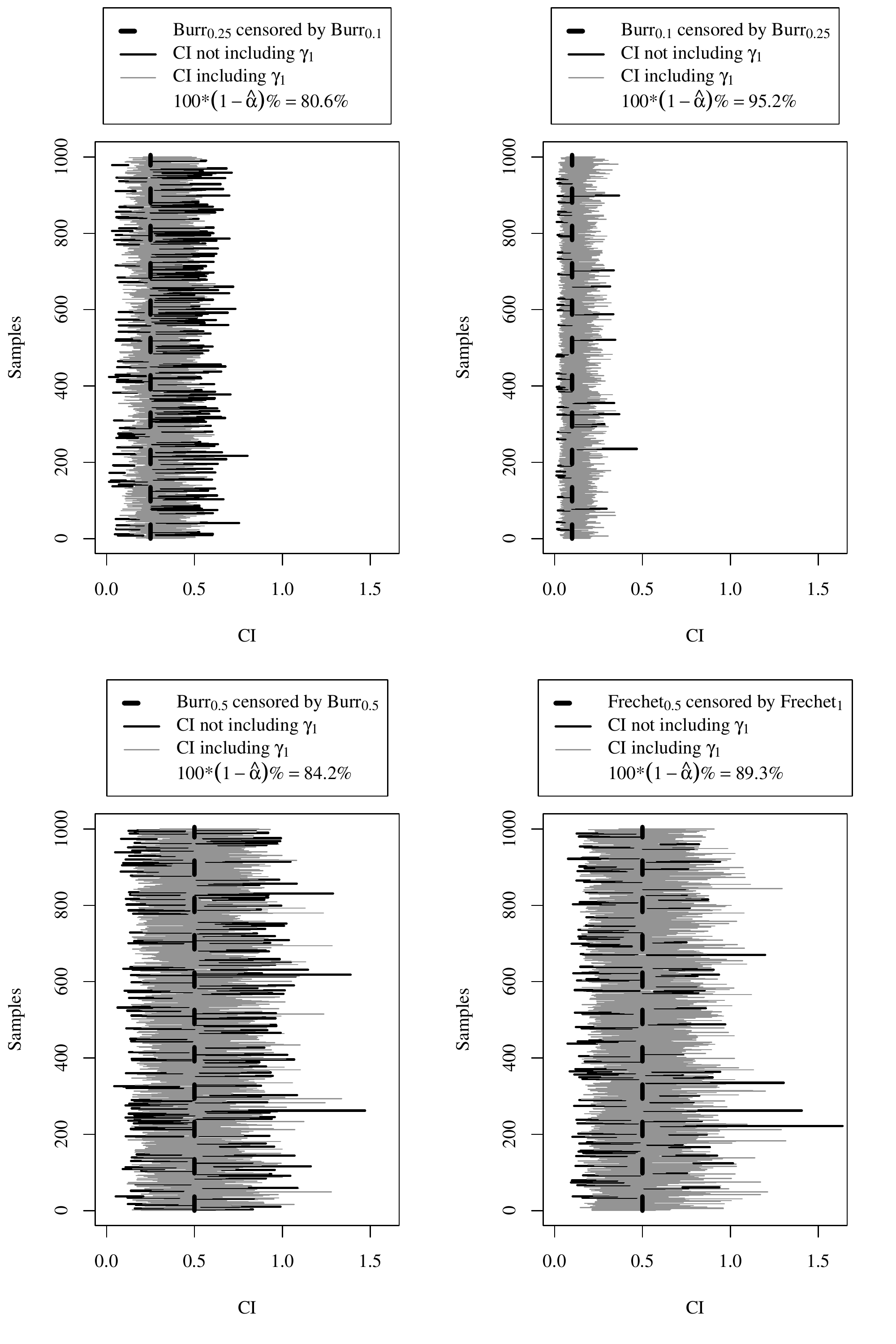}
		\includegraphics[width=0.6\textwidth]{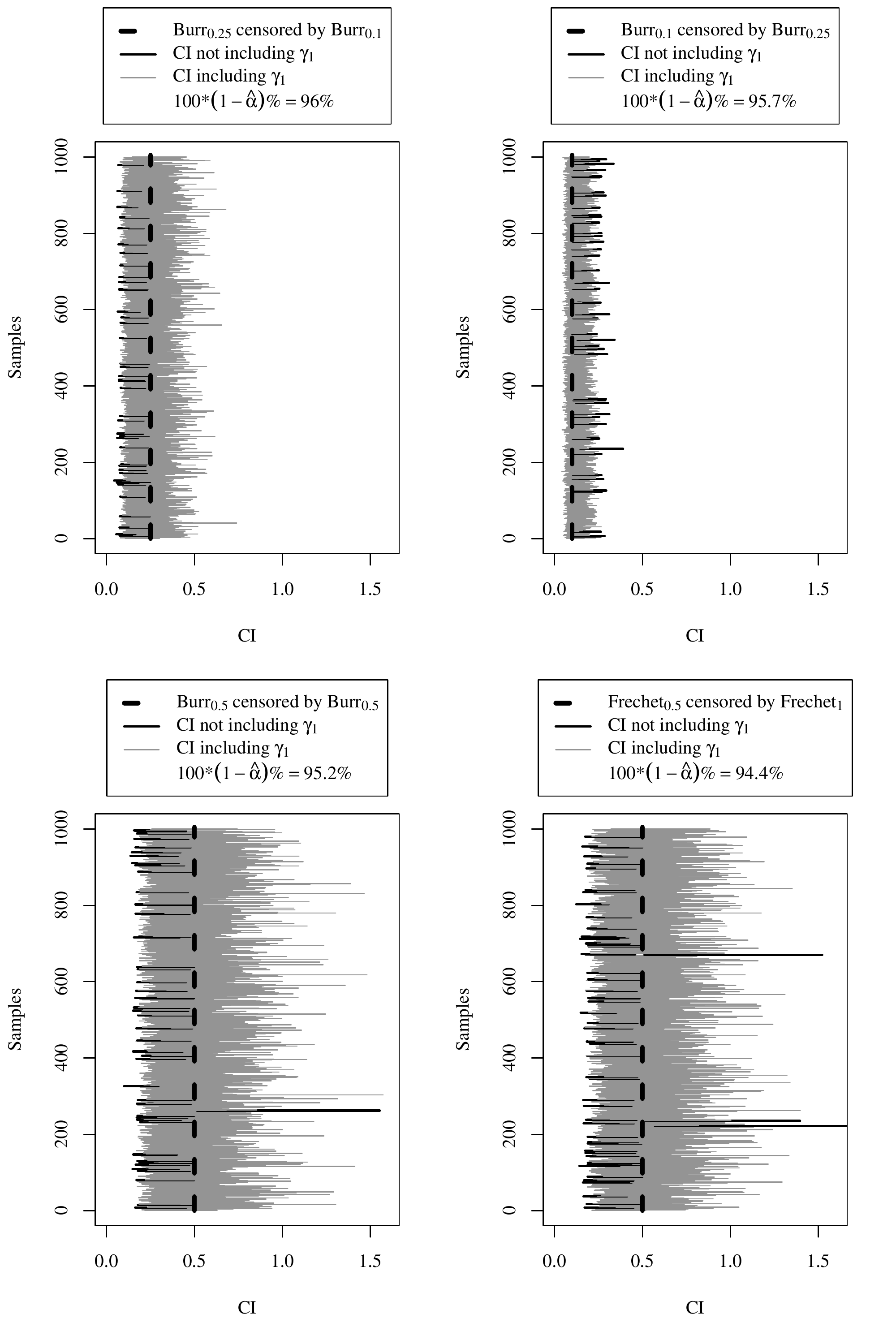}
		\caption{Simulated bootstrap 95\% confidence intervals with adaptive choice of $k_1$ and $k_2$ using $\epsilon = 0.01$ (4 left frames) and using $k_1=k_2=25=5\%n$ for every sample (4 right frames)}
			\end{figure}
\end{landscape}

\begin{figure}[h]
		\centering
		\includegraphics[width=0.7\textwidth]{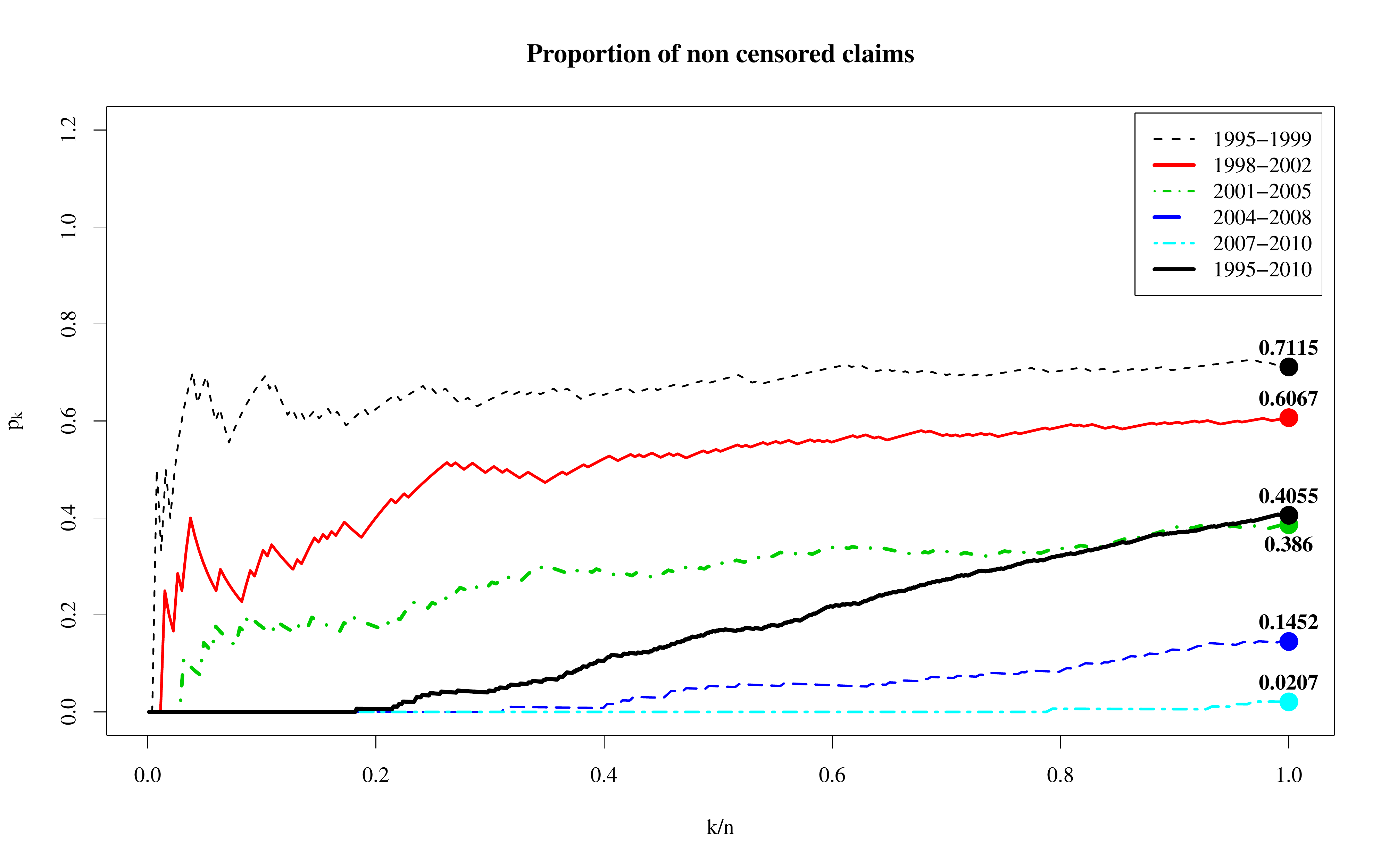}
		\caption{Car liability data: plot of $\hat{p}_k$ as a function of $k/n$ for all claims jointly (thick full line) and for claims that arrived in different time intervals.} 
	\end{figure}

\begin{figure}[h]
		\centering
		\includegraphics[width=0.49\textwidth]{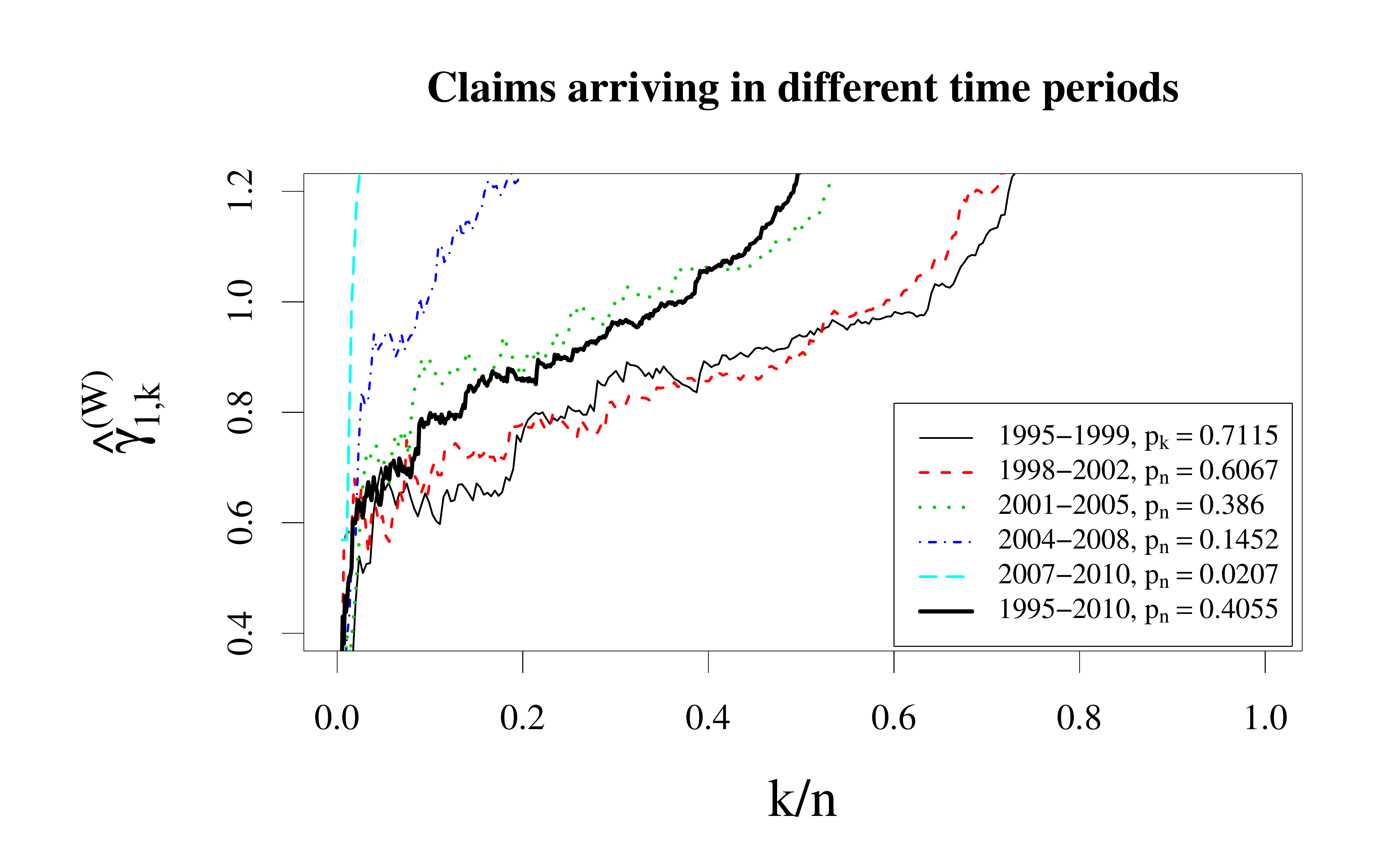}
	\includegraphics[width=0.49\textwidth]{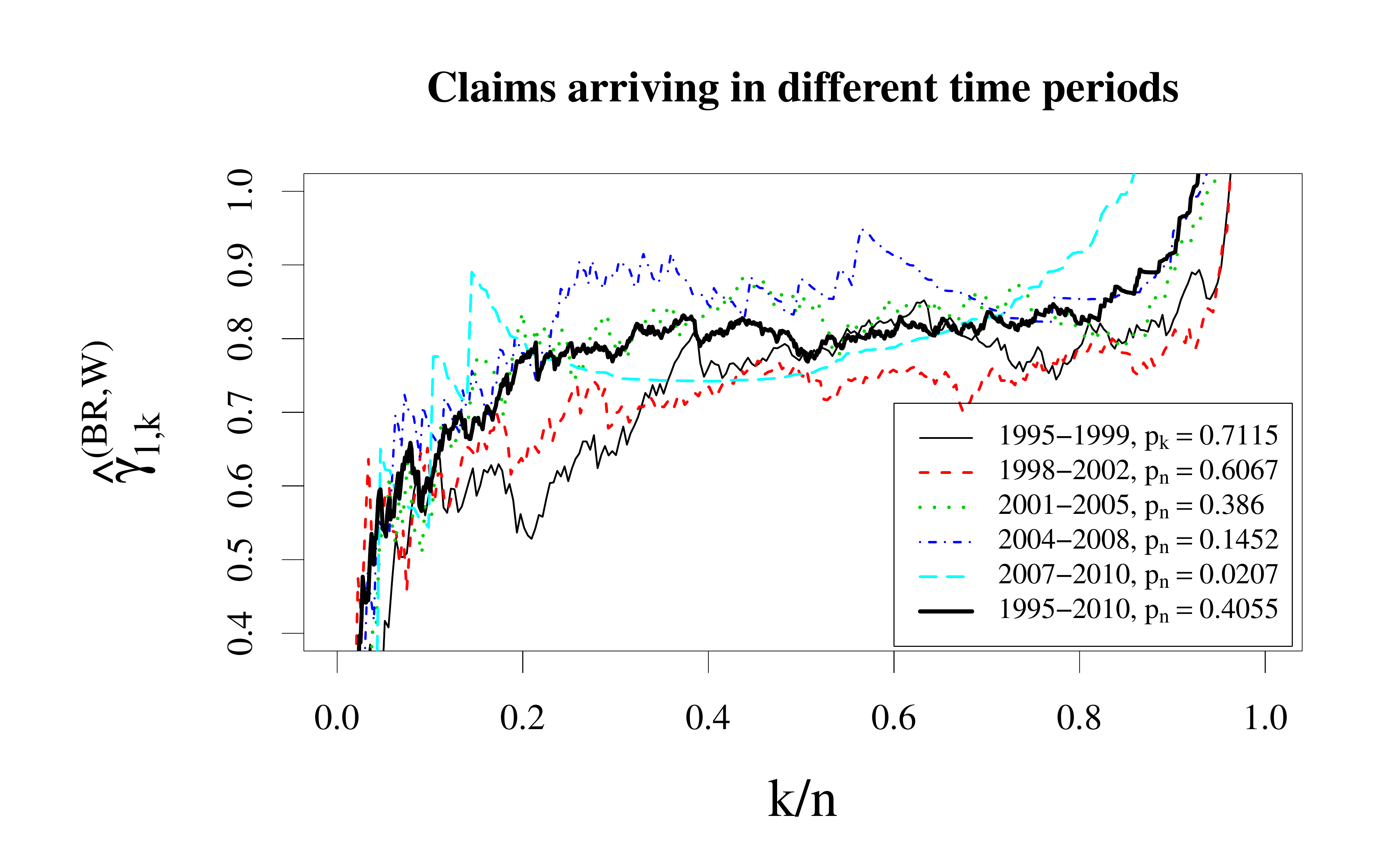}
		\caption{Car liability data: $\hat{\gamma}_{1,k}^{(W)}$ (left) and  $\hat{\gamma}_{1,k}^{(BR,W)}(-3)$ (right) based on indexed cumulative payments at end of 2010, as a function of $k/n$ for all claims together (thick full line) and for claims arriving in different time periods}
	\end{figure}
	
	\begin{figure}[h]
		\centering
		\includegraphics[width=0.49\textwidth]{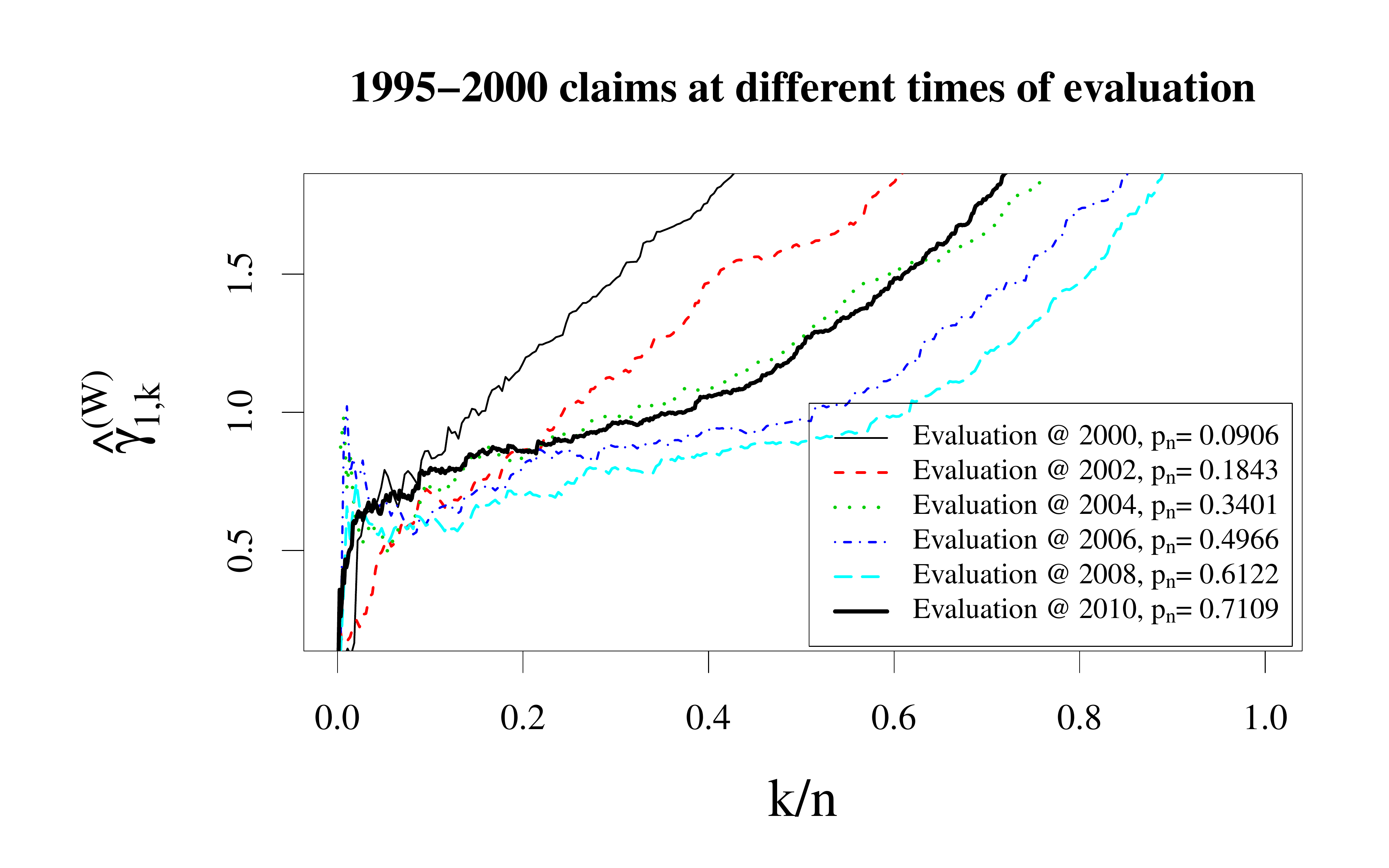}
	\includegraphics[width=0.49\textwidth]{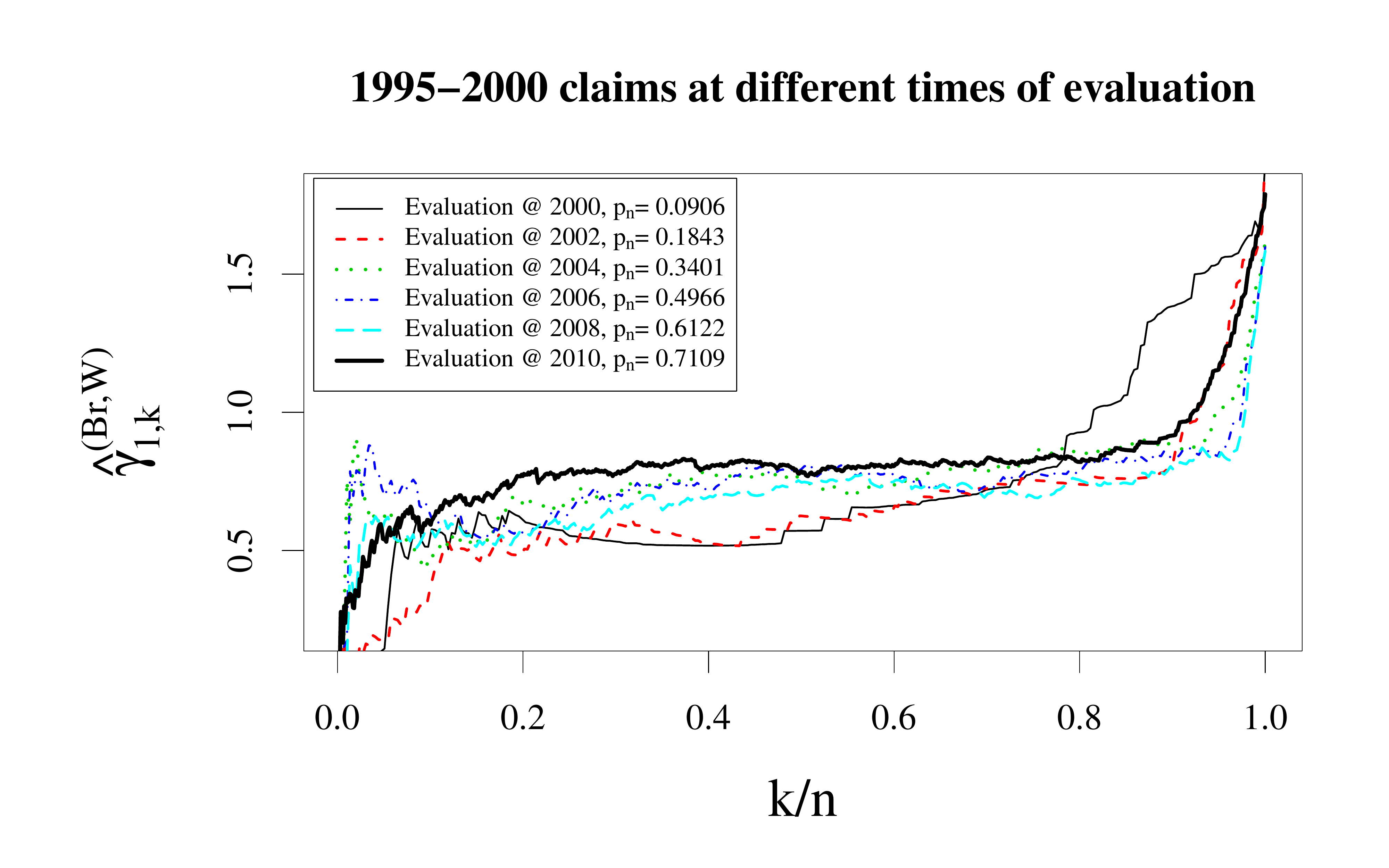}
		\caption{Car liability data: $\hat{\gamma}_{1,k}^{(W)}$ (left) and  $\hat{\gamma}_{1,k}^{(BR,W)}(-3)$ (right) for all claims arriving in 1995-1999 and  based on indexed cumulative payments at end of 2000(02)2010, as a function of $k/n$.  }
	\end{figure}
	
	\begin{figure}[h]
		\centering
			\includegraphics[width=0.49\textwidth]{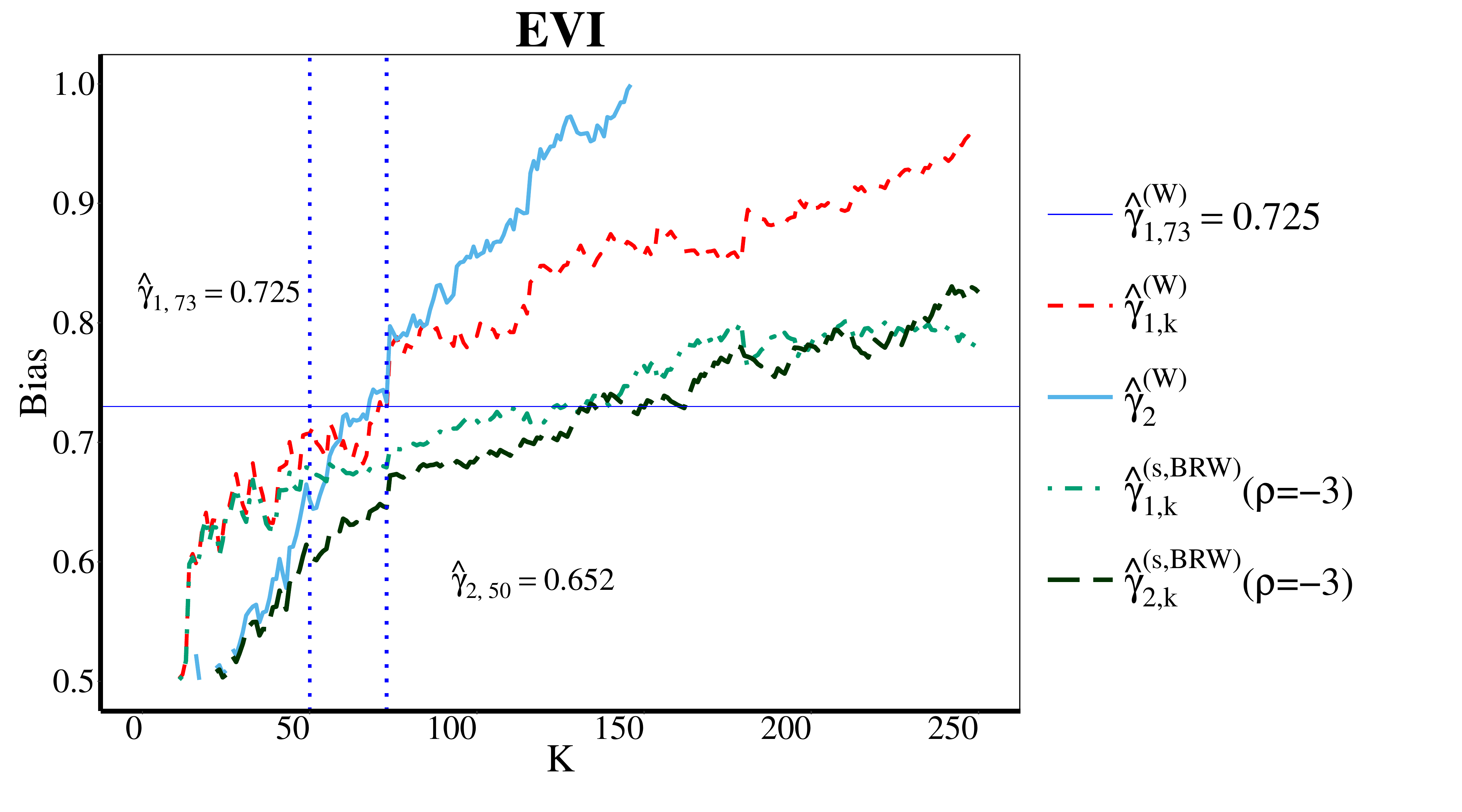}
			\includegraphics[width=0.49\textwidth]{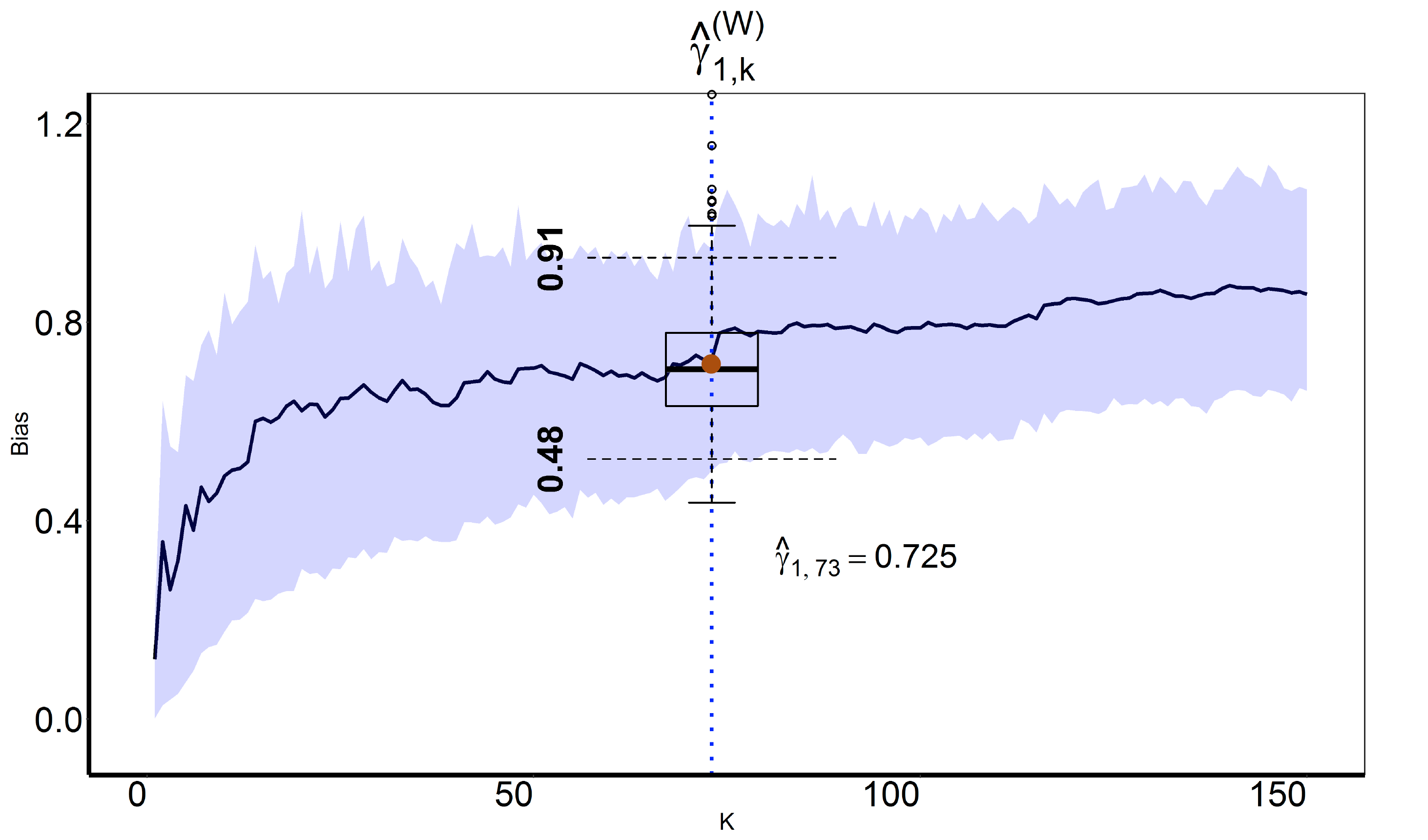}
		\caption{Car liability data, using all claims: $\hat{\gamma}_{1,k}^{(W)}$, $\hat{\gamma}_{1,k}^{(s,W)}(-3)$, $\hat{\gamma}_{2,k}^{(W)}$, $\hat{\gamma}_{2,k}^{(s,W)}(-3)$ (left); 95\% bootstrap confidence intervals (right).}
	\end{figure}
\end{document}